\documentclass[%
 reprint,
 amsmath,amssymb,
 aps,
pra,
]{revtex4-1}

\usepackage{subfigure}
\usepackage{graphicx}
\usepackage{dcolumn}
\usepackage{bm}
\usepackage{amsmath,amsfonts,amssymb}

\begin{document}


\title{Resonant shortcuts for adiabatic rapid passage with only z-field control}

\author{Dionisis Stefanatos}
\email{dionisis@post.harvard.edu}
\author{Emmanuel Paspalakis}%
\affiliation{Materials Science Department, School of Natural Sciences, University of Patras, Patras 26504, Greece}

\date{\today}

\begin{abstract}

In this work we derive novel ultrafast shortcuts for adiabatic rapid passage for a qubit where the only control variable is the longitudinal $z$-field, while the transverse $x$-field remains constant. This restrictive framework is pertinent to some important tasks in quantum computing, for example the design of a high fidelity controlled-phase gate can be mapped to the adiabatic quantum control of such a qubit. We study this problem in the adiabatic reference frame and with appropriately rescaled time, using as control input the derivative of the total field polar angle (with respect to rescaled time). We first show that a constant pulse can achieve perfect adiabatic rapid passage at only specific times, corresponding to resonant shortcuts to adiabaticity. We next show that, by using ``on-off-on-...-on-off-on" pulse-sequences with appropriate characteristics (amplitude, timing, and number of pulses), a perfect fidelity can be obtained for every duration larger than the limit $\pi/\Omega$, where $\Omega$ is the constant transverse $x$-field. We provide equations from which these characteristics can be calculated. The proposed methodology for generalized resonant shortcuts exploits the advantages of composite pulses in the rescaled time, while the corresponding control $z$-field varies continuously and monotonically in the original time. Of course, as the total duration approaches the lower limit, the changes in the control signal become more abrupt. These results are not restricted only to quantum information processing applications, but is also expected to impact other areas, where adiabatic rapid passage is used.

\end{abstract}

\maketitle

\section{Introduction}

\label{sec:intro}

A central problem in modern quantum technology is the efficient control of two-level quantum systems \cite{Roadmap17,Glaser15}. Adiabatic rapid passage (ARP) \cite{Vitanov01,Goswami03} is one of the most widely used quantum control techniques to tackle this problem \cite{Glaser15}. The method has been proven to be robust to moderate variations of the system parameters. The major limitation of this technique, as with every adiabatic method, is the necessary long operation time, which may lead to a degraded performance in the presence of decoherence and dissipation.

In order to accelerate adiabatic quantum dynamics, several methods have been suggested over the years, collectively characterized as \emph{Shortcuts to Adiabaticity} \cite{Demirplak03,Berry09,Motzoi09,Chen10a,Masuda10,Deffner14}. The basic concept behind this technique is to drive the quantum system at the same final state as with a slow adiabatic process, but without necessarily following the instantaneous adiabatic eigenstates at intermediate times. These methods have been used in
a broad spectrum of applications including two-level quantum systems \cite{Chen10,Chen11,Bason12,Malossi13,Ruschhaupt12,Daems13,Ibanez13,Motzoi13,Theis18}, where both the longitudinal ($z$) and transverse ($x$) fields are exploited in order to speed up adiabatic evolution.

Although the general two-level system framework, where both fields can serve as control inputs, covers a wide range of applications, it turns out that for several important core tasks in quantum computing a more restrictive framework is pertinent, where only the $z$-field is time-dependent while the $x$-field is constant, see Refs. \cite{Martinis14a,Martinis14,Shim16,Zeng18NJP,Zeng18PRA,Fischer19}. As a specific example we mention the design of a high fidelity controlled-phase gate \cite{Martinis14a}, which can be mapped to the adiabatic quantum control of such a qubit \cite{Martinis14}. The authors of \cite{Martinis14} study this problem in the adiabatic reference frame and with an appropriate rescaling of time. They use as control variable the instantaneous polar angle of the total field or its derivative with respect to the rescaled time, and express them as Fourier series with few components, which also satisfy appropriate boundary conditions at the beginning and at the end of the applied time interval. Finally, they obtain numerically the coefficients in the series which minimize the error of the final state with respect to the adiabatic evolution. They find that, even for short durations of the control function (a few times the system timescale) high fidelity is achieved, at levels appropriate for fault-tolerant quantum computation.

In the present article we also study the adiabatic quantum control of a qubit where only the $z$-field can vary in time, while the $x$-field is fixed. Following \cite{Martinis14}, we work in the adiabatic reference frame and use as control variable the derivative with respect to rescaled time of the field polar angle. We start with a constant control pulse (in rescaled time) and show that a perfect fidelity can be obtained for specific durations, a phenomenon also observed for certain nonlinear Landau-Zener sweeps \cite{Garanin02}. The error probability becomes zero for these durations, corresponding to a series of resonant shortcuts to adiabaticity. After determining the time dependence of the $z$-field in the original time, it becomes obvious that the constant control pulse (in rescaled time) producing these shortcuts is exactly the Roland-Cerf adiabatic protocol \cite{Roland02}, considered at specific times. Next, we consider more general control inputs, specifically pulse-sequences of the form ``on-off-on-...-on-off-on". We explain that by appropriately choosing the pulse-sequence characteristics (the amplitude, the timing and the number of the pulses), a perfect fidelity can be obtained for any total duration $T\geq\pi/\Omega$, where $\Omega$ is the constant transverse $x$-field. We provide equations from which the amplitude of the ``on" segments and the durations of both the ``on" and ``off" segments can be calculated for any total duration above this limit. The number of pulses in the sequence is determined in a systematic way that we also explain. The suggested methodology takes advantage of composite pulse characteristics \cite{Levitt86,Torosov11,Torosov18,Torosov19} in the rescaled time, while the corresponding polar control angle varies continuously and monotonically in the original time, as well as the control $z$-field. These characteristics differentiate the present study from previous works where the longitudinal field is also the solely control but it changes discontinuously and non-monotonically \cite{Hegerfeldt13,Hegerfeldt14}. The resultant generalized resonant shortcuts can implement ARP with any desired duration larger than the limit specified above. Of course, as the lower bound for the total duration is approached, the corresponding control signal varies more abruptly. The application of the present work is not restricted only to quantum information processing, but may also span other areas where ARP is used, like nuclear magnetic resonance and spectroscopy.

The structure of the paper is as follows. In the next section we derive the resonant shortcuts resulting from a constant control pulse. In Section \ref{sec:generalized} we describe the methodology for obtaining generalized resonant shortcuts for every permitted duration using pulse-sequences as control inputs, while in Section \ref{sec:examples} we clarify the procedure with several examples. Section \ref{sec:conclusion} concludes this work.

\section{Resonant technique for adiabatic rapid passage}

\label{sec:resonant}

We consider a two-level system with Hamiltonian
\begin{equation}
H(t)=\frac{\Delta(t)}{2}\sigma_z+\frac{\Omega}{2}\sigma_x=\frac{1}{2}
\left[
\begin{array}{cc}
\Delta(t) & \Omega\\
\Omega & -\Delta(t)
\end{array}
\right], \label{Hamiltonian}
\end{equation}
where $\sigma_x, \sigma_z$ are the Pauli spin matrices. The Rabi frequency $\Omega$ ($x$-field) is constant while the time-dependent detuning $\Delta(t)$ ($z$-field) is the control parameter. The instantaneous angle $\theta$ of the total field with respect to $z$-axis is
\begin{equation}
\label{theta}
\cot{\theta(t)}=\frac{\Delta(t)}{\Omega},
\end{equation}
and can also serve as a control parameter instead of the detuning. In terms of $\theta$, Hamiltonian (\ref{Hamiltonian}) is expressed as
\begin{equation}
\label{parametrized_H}
H=\frac{\Omega}{2\sin{\theta}}
\left(\begin{array}{cc}
    \cos\theta & \sin\theta \\ \sin\theta & -\cos\theta
  \end{array}\right).
\end{equation}
If $|\psi\rangle=a_1|0\rangle+a_2|1\rangle$ denotes the state of the system, the probability amplitudes $\mathbf{a}=(a_1 \; a_2)^T$ obey the equation ($\hbar=1$)
\begin{equation}
\label{Schrodinger_a}
i\dot{\mathbf{a}}=H\mathbf{a}.
\end{equation}
The normalized eigenvectors of Hamiltonian (\ref{parametrized_H}) are
\begin{subequations}
\label{eigenvectors}
\begin{eqnarray}
|\phi_{+}(t)\rangle&=&
\left(\begin{array}{c}
    \cos{\frac{\theta}{2}}\\
    \sin{\frac{\theta}{2}}
\end{array}\right),\label{plus}\\
|\phi_{-}(t)\rangle&=&
\left(\begin{array}{c}
    \sin\frac{\theta}{2}\\
    -\cos{\frac{\theta}{2}}
\end{array}\right).\label{minus}
\end{eqnarray}
\end{subequations}
In the traditional ARP, we start at $t=0$ with a large negative value of the detuning, $-\Delta(0)$ with $|\Delta(0)|\gg\Omega$, so the initial angle $\theta(0)=\theta_i\approx \pi$. Therefore, $|\phi_{+}(0)\rangle = (0 \; 1)^T$ and $|\phi_{-}(0)\rangle = (1 \; 0)^T$.  Then, the detuning is increased linearly with time, until it reaches a large positive value $\Delta(T)=|\Delta(T)|\gg\Omega$ at the final time $t=T$. If the change is slow enough, i.e. for a sufficiently long duration $T$, the system remains in the same eigenstate of the instantaneous Hamiltonian. For large final positive detuning it is $\theta(T)=\theta_f\approx 0$. So, the final states are close to $|\phi_{+}(T)\rangle = (1 \; 0)^T$ and $|\phi_{-}(T)\rangle = (0 \; -1)^T$. As at early and final times each adiabatic state becomes uniquely identified with one of the original states of the system, ARP achieves complete population transfer from state $|0\rangle$ to $|1\rangle$ and vice versa. It is a well known fact that the method is robust to moderate variation of the system parameters. The drawback of the method is the long necessary time which might be crucial in the presence of decoherence and dissipation. In this and the next sections we derive controls, $\Delta(t)$ and $\theta(t)$, which implement shortcuts of the adiabatic evolution by driving the system to the same final eigenstate without following the intermediate adiabatic path.

In order to obtain these shortcuts it is more convenient to work in the adiabatic frame.
By expressing the state of the system in both the original and the adiabatic frames
\begin{equation}
\label{adiabatic_basis}
|\psi\rangle=a_1|0\rangle+a_2|1\rangle=b_1|\phi_{+}\rangle+b_2|\phi_{-}\rangle,
\end{equation}
we obtain the following transformation between the probability amplitudes of the two pictures
\begin{equation}
\label{transformation}
\mathbf{b}=
\left(\begin{array}{c}
    b_1\\
    b_2
\end{array}\right)
=
\left(\begin{array}{cc}
    \cos{\frac{\theta}{2}} & \sin{\frac{\theta}{2}} \\ \sin{\frac{\theta}{2}} & -\cos{\frac{\theta}{2}}
  \end{array}\right)
\left(\begin{array}{c}
    a_1\\
    a_2
\end{array}\right).
\end{equation}
From Eqs. (\ref{Schrodinger_a}), (\ref{transformation}) we find the following equation for the probability amplitudes in the adiabatic frame
\begin{equation}
\label{Schrodinger_b}
i\dot{\mathbf{b}}=H_{ad}\mathbf{b},
\end{equation}
where the Hamiltonian now is
\begin{equation}\label{adiabatic_H}
H_{ad}=\frac{1}{2}
\left(\begin{array}{cc}
    \frac{\Omega}{\sin{\theta}} & -i\dot{\theta} \\ i\dot{\theta} & -\frac{\Omega}{\sin{\theta}}
  \end{array}\right).
\end{equation}

The next step is to use a dimensionless rescaled time $\tau$, related to the original time $t$ through the relation \cite{Martinis14}
\begin{equation}
\label{rescaled_time}
d\tau=\frac{\Omega}{\sin{\theta}}dt.
\end{equation}
For $0<\theta<\pi$ that we consider here it is $\sin{\theta}>0$ and the rescaling (\ref{rescaled_time}) is well defined.
The equation for $\mathbf{b}$ becomes
\begin{equation}
\label{rescaled_Schrodinger_b}
i\mathbf{b}'=H'_{ad}\mathbf{b},
\end{equation}
where
\begin{equation}
\label{rescaled_adiabatic_H}
H'_{ad}=\frac{1}{2}\sigma_z+\frac{\theta'}{2}\sigma_y \, ,
\end{equation}
and $\mathbf{b}'=d\mathbf{b}/d\tau$, $\theta'=d\theta/d\tau$ are the derivatives with respect to the rescaled time.
Here we consider a general change in the angle $\theta$, from some initial value $\theta_i$ at $\tau=0$ to some final value $\theta_f$ at $\tau=T'$ (we use prime to denote the duration in rescaled time), with $\theta_i>\theta_f$, and we set
\begin{equation}
\label{control}
\theta'=\frac{d\theta}{d\tau}=-u<0.
\end{equation}
Now observe that for \emph{constant} derivative $u$ the Hamiltonian $H'_{ad}$ is also constant and from Eq. (\ref{rescaled_Schrodinger_b}) we obtain $\mathbf{b}(T')=U\mathbf{b}(0)$, where the unitary transformation $U$ is given by
\begin{eqnarray}
\label{U}
U&=&e^{-iH'_{ad}T'}\nonumber\\
&=&e^{-i\frac{1}{2}\omega T'(n_z\sigma_z-n_y\sigma_y)}\nonumber\\
&=& I\cos{\frac{\omega T'}{2}}-i\sin{\frac{\omega T'}{2}}(n_z\sigma_z-n_y\sigma_y) \, ,
\end{eqnarray}
and
\begin{eqnarray*}
\label{parameters}
\omega&=&\sqrt{1+u^2},\\
n_z&=&\frac{1}{\omega}=\frac{1}{\sqrt{1+u^2}},\\
n_y&=&\frac{u}{\omega}=\frac{u}{\sqrt{1+u^2}}.
\end{eqnarray*}

The system starts in the $|\phi_{+}\rangle$ state, thus $\mathbf{b}(0)=(1 \; 0)^T$. In order to implement a shortcut to adiabaticity the system should end up in the same state at $\tau=T'$, thus it is sufficient that $b_2(T')=0$. From Eq. (\ref{U}) we obtain the condition $\sin{(\omega T'/2)}=0$, such that $U=\pm I$, which leads to
\begin{equation}
\label{shortcut_evolution}
T'\sqrt{1+u^2}=2k\pi,\quad k=1,2,\ldots
\end{equation}
During the time $T'$ the angle should change from $\theta_i$ to $\theta_f$, thus
\begin{equation}
\label{angle_evolution}
\theta_i-\theta_f=-\int_0^{T'}\theta'd\tau=uT' \, ,
\end{equation}
for constant $u$.
Combining Eqs. (\ref{shortcut_evolution}) and (\ref{angle_evolution}) we find the solution pairs
\begin{subequations}
\begin{eqnarray}
u_k&=&\frac{\frac{\theta_i-\theta_f}{2k\pi}}{\sqrt{1-\left(\frac{\theta_i-\theta_f}{2k\pi}\right)^2}},\label{uk}\\
T'_k&=&2k\pi\sqrt{1-\left(\frac{\theta_i-\theta_f}{2k\pi}\right)^2},\label{Tk}
\end{eqnarray}
\end{subequations}
for $k=1,2,\ldots$.

\begin{figure*}[t]
 \centering
		\begin{tabular}{cc}
     	\subfigure[$\ $]{
	            \label{fig:pulse0}
	            \includegraphics[width=.45\linewidth]{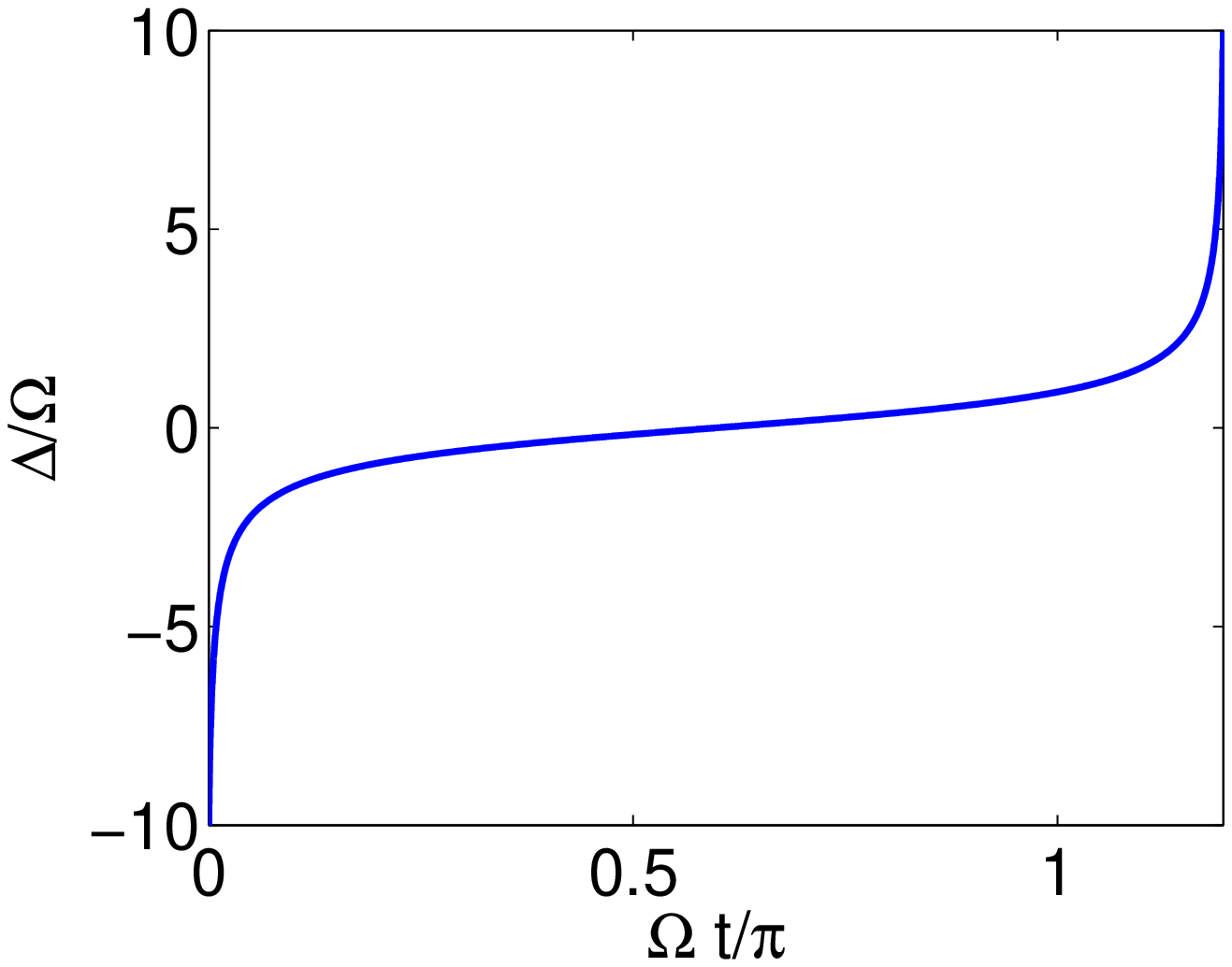}} &
       \subfigure[$\ $]{
	            \label{fig:theta0}
	            \includegraphics[width=.45\linewidth]{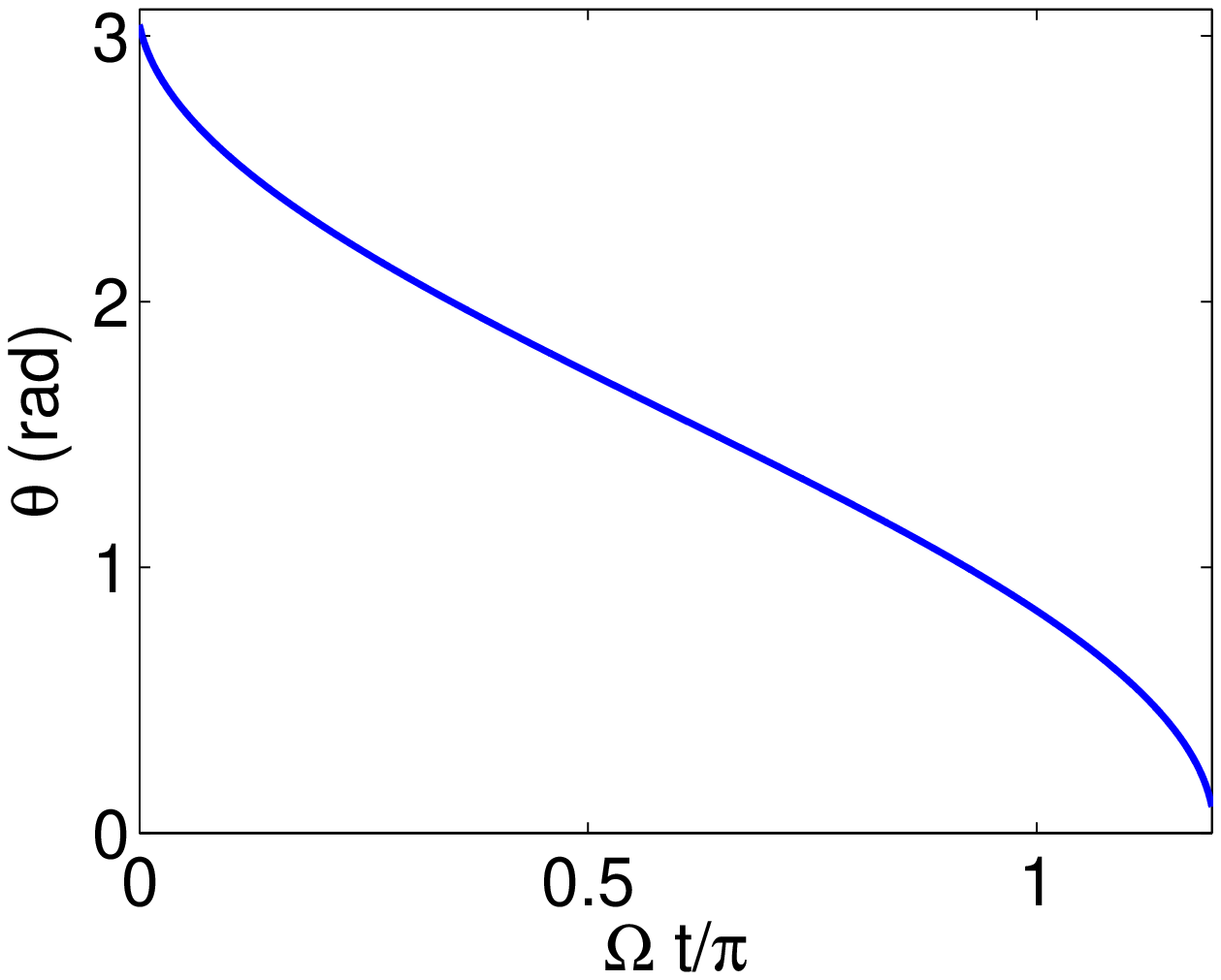}} \\
      \subfigure[$\ $]{
	            \label{fig:original0}
	            \includegraphics[width=.45\linewidth]{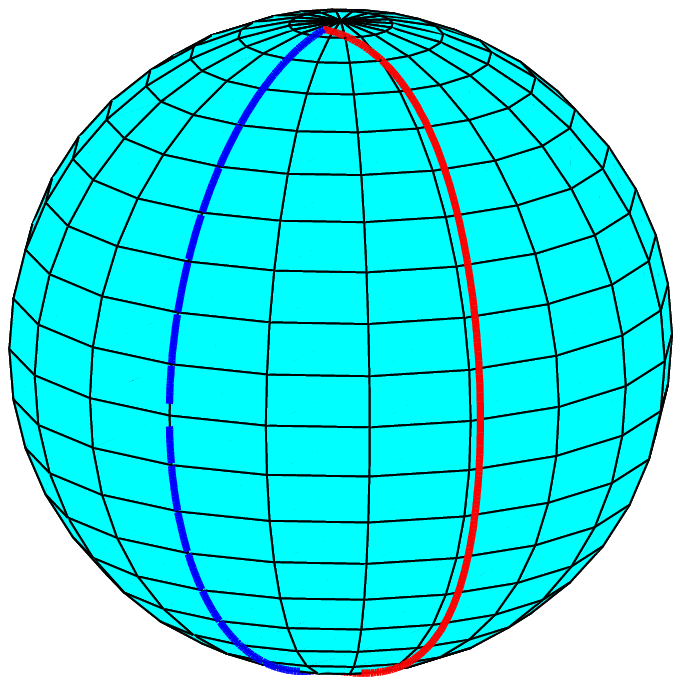}} &
       \subfigure[$\ $]{
	            \label{fig:adiabatic0}
	            \includegraphics[width=.45\linewidth]{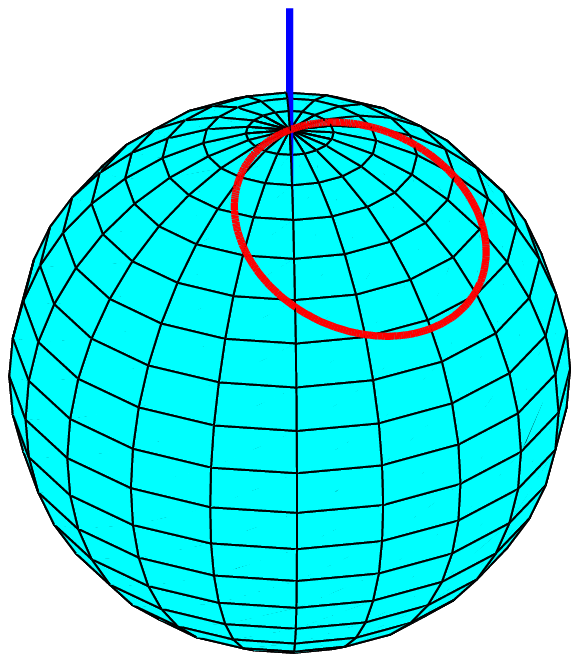}}
		\end{tabular}
\caption{(Color online) (a) Detuning $\Delta(t)$ corresponding to the first resonance $k=1$ with duration $T_1\approx 1.195\pi/\Omega$. (b) Corresponding evolution of the total field angle $\theta(t)$, also in the original time $t$. (c) State trajectory (red solid line) on the Bloch sphere in the original reference frame. The blue solid line on the meridian lying on the $xz$-plane indicates the change in the total field angle $\theta$. (d) State trajectory (red solid line) on the Bloch sphere in the adiabatic frame. Observe that in this frame the state of the system returns to the north pole, while the total field points constantly in the $\hat{z}$-direction (blue solid line).}
\label{fig:constant}
\end{figure*}

\begin{figure}[t]
\centering
\includegraphics[width=0.7\linewidth]{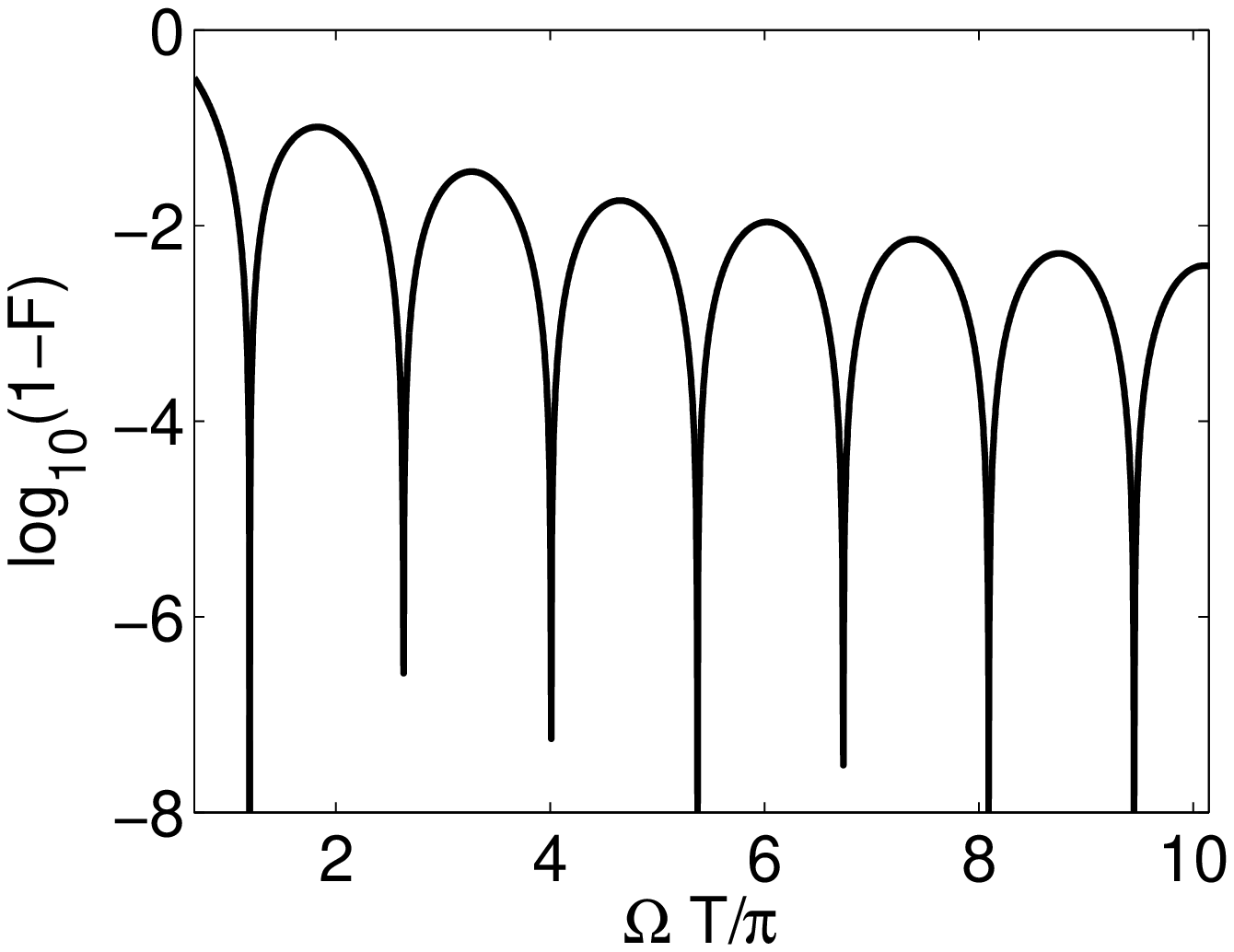}
\caption{Logarithmic error for a constant pulse $u$ (in the rescaled time $\tau$), as a function of the duration $T$ in the original time $t$. The resonances corresponding to durations $T=T_k$ given in Eq. (\ref{durations_original}) are clearly displayed.}
\label{fig:constant_fidelity}
\end{figure}

Angle $\theta$ varies linearly with respect to the rescaled time $\tau$
\begin{equation}
\label{angle_rescaled}
\theta_k(\tau)=\theta_i-u_k\tau,
\end{equation}
while from Eq. (\ref{rescaled_time}) we can easily obtain its dependence on the original time $t$
\begin{equation}
\label{angle_original}
\theta_k(t)=\cos^{-1}(\cos{\theta_i}+u_k\Omega t).
\end{equation}
Note that we have parameterized these solutions with the positive integer $k$ used in Eqs. (\ref{uk}), (\ref{Tk}) .
The durations corresponding to $T'_k$ in the original time $t$ are
\begin{equation}
\label{durations_original}
T_k=\frac{\cos{\theta_f}-\cos{\theta_i}}{u_k}\cdot\frac{1}{\Omega}.
\end{equation}
Using Eqs. (\ref{theta}), (\ref{angle_original}) we can also find the corresponding detuning $\Delta(t)$. In the symmetric case $\theta_f=\pi-\theta_i$, where $\cos{\theta_f}=-\cos{\theta_i}$, if we use a \emph{shifted} time $t_s=t-T_k/2$, so $\theta_k(t_s)=\cos^{-1}(u_k\Omega t_s)$ and $\theta_k(t_s=0)=\pi/2$, we obtain the following simple expression
\begin{equation}
\label{Delta}
\Delta(t_s)=\frac{u\Omega t_s}{\sqrt{1-(u\Omega t_s)^2}}\Omega,
\end{equation}
for $-T_k/2\leq t_s\leq T_k/2$. This detuning form is exactly the one derived from the Roland-Cerf adiabatic protocol \cite{Roland02}, see for example Refs. \cite{Bason12,Malossi13}, thus it turns out that the resonant shortcuts presented above actually implement this protocol for the specific durations (\ref{durations_original}).

As an example we consider a change in the detuning $\Delta$ from $-10\Omega$ to $10\Omega$, same as in \cite{Martinis14}, corresponding to $\theta_f=\tan^{-1}(1/10)$, $\theta_i=\pi-\theta_f$. In Fig. \ref{fig:pulse0} we plot the detuning $\Delta(t)$ corresponding to the duration $T_1\approx 1.195\pi/\Omega$ of the first resonance. In Fig. \ref{fig:theta0} we show the corresponding evolution of the total field angle $\theta(t)$ in the original time $t$. In Fig. \ref{fig:original0} we plot with red solid line the corresponding state trajectory on the Bloch sphere and in the original reference frame. The blue solid line on the meridian indicates the change in the total field angle $\theta$. Finally, in Fig. \ref{fig:adiabatic0} we plot the same trajectory (red solid line) but in the adiabatic frame. Note that in this frame the system starts from the adiabatic state at the north pole and returns there at the final time, while the total field points constantly in the $\hat{z}$-direction (blue solid line). In the next Fig. \ref{fig:constant_fidelity}, we show the logarithmic error at the final time
\begin{equation}
\label{logarithmic_error}
\log_{10}{(1-F)}=\log_{10}{|b_2(T')|^2} \, ,
\end{equation}
for a constant pulse (in the rescaled time)
\begin{equation}
u=\frac{\theta_i-\theta_f}{T'},
\end{equation}
as a function of the duration $T$ in the original time $t$ (corresponding to $T'$ in the rescaled time $\tau$). The resonances for durations $T=T_k$ given in Eq. (\ref{durations_original}) are clearly displayed.

Similar resonant shortcuts to adiabaticity have been derived for quantum teleportation \cite{Oh13} but using two control fields instead of the one that we have here, which actually play the role of Stokes and pump pulses in the familiar STIRAP terminology \cite{Kral07,Vitanov17}.
Resonant shortcuts have also been obtained for the quantum parametric oscillator, see for example \cite{Rezek09,Kosloff17}. In the subsequent section we generalize the above analysis and obtain shortcuts with arbitrary duration
\begin{equation}
\label{lower_bound}
T>T_0=\sin{\frac{\theta_i+\theta_f}{2}}\frac{\pi}{\Omega}\Rightarrow T'>T'_0=\pi,
\end{equation}
where in the second inequality the bound is expressed in rescaled time. Before moving to the next section, we briefly explain how this lower bound in the duration is obtained. In the \emph{original reference frame} (not the adiabatic), we consider an instantaneous change in the total field from $\theta=\theta_i$ to $\theta=\bar{\theta}=(\theta_i+\theta_f)/2$, i.e. in the middle of the arc connecting the initial and target states. The corresponding detuning is $\Delta=\Omega\cot{\bar{\theta}}$ and the total field is $\sqrt{\Delta^2+\Omega^2}=\Omega/\sin{\bar{\theta}}$. Under the influence of this constant field for duration $T_0=\sin{\bar{\theta}}\cdot\pi/\Omega$, the Bloch vector is rotated from $(\phi=0,\theta_i)$ to $(\phi=0,\theta_f)$. After the completion of this half circle, the total field is changed again instantaneously from $\theta=\bar{\theta}$ to $\theta=\theta_f$. Since $\theta=\bar{\theta}$ during this evolution, except the (measure zero) initial and final instants, Eq. (\ref{rescaled_time}) becomes $d\tau=\Omega dt/\sin{\bar{\theta}}$ and the corresponding duration in the rescaled time is thus $T'_0=\Omega T_0/\sin{\bar{\theta}}=\pi$.
We finally point out that the corresponding quantum speed limit (in the original time) is $T_{qsl}=(\theta_i-\theta_f)/\Omega$, as derived in \cite{Hegerfeldt13} and also mentioned in \cite{Poggi13}, but is obtained using infinite values of the detuning which implement instantaneous rotations around $z$-axis, while angle $\theta$ changes non-monotonically. More realistic speed limits have been obtained for bounded detuning \cite{Hegerfeldt14}, but their implementation also requires discontinuous and non-monotonic changes of the magnetic field angle. On the contrary, the bounds in Eq. (\ref{lower_bound}) are obtained with finite detuning values and a monotonic change of $\theta$ (decrease for $\theta_i>\theta_f$). For $\theta_i\approx\pi$ and $\theta_f\approx 0$, it is $T_{qsl}\approx T_0\approx \pi/\Omega$, as derived in \cite{Boscain02}.


\section{Generalized Resonant Shortcuts}

\label{sec:generalized}

\begin{figure}[t]
\centering
\includegraphics[width=0.7\linewidth]{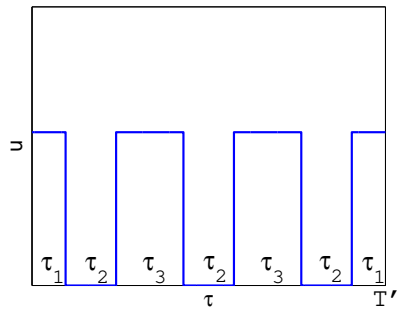}
\caption{Representative example of the pulse-sequences $u(\tau)$ in the rescaled time $\tau$ that we consider in this section. The initial and final ``on" pulses have the same duration $\tau_1$, all the intermediate ``off" pulses have the same duration $\tau_2$, while all the intermediate ``on" pulses have the same duration $\tau_3$. This timing of the pulses is motivated by optimal control theory and the symmetry of the problem. The middle pulse can be ``off", as in this figure, or ``on". The total duration of the sequence is denoted by $T'$.}
\label{fig:pulse_sequence}
\end{figure}

In the previous section we derived resonant shortcuts using a constant control $u=-d\theta/d\tau$ (in rescaled time $\tau$). In order to obtain shortcuts with arbitrary duration, the idea is to use a time-dependent $u(\tau)$ but with a simple on-off modulation in the rescaled time $\tau$. We will consider pulse-sequences $u(\tau)$ of the form ``on-off-on-...on-off-on", see Fig. \ref{fig:pulse_sequence}, with the following characteristics: all the ``on" pulses have the same constant amplitude $u$, the first and the last ``on" pulses have the same duration $\tau_1$, all the intermediate ``off" pulses have the same duration $\tau_2$, while all the intermediate ``on" pulses have the same duration $\tau_3$. The equality between the durations of corresponding intermediate pulses is motivated from time-optimal control theory, see for example our previous works \cite{Stefanatos11,Stefanatos14,Stefanatos2017a,Stefanatos2017b,Stefanatos_PRE17} where similar optimal pulse-sequences are derived. The equality between the durations of the initial and final pulses comes from the symmetry of the problem which is apparent in the adiabatic frame, where the initial and final (target) states coincide on the Bloch sphere with the north pole. With the appropriate choice of the pulse amplitude $u$, the durations $\tau_i$, $i=1,2,3$, and the number of pulses, we can assure that the system returns at the final time $\tau=T'$ to the adiabatic state $|\phi_{+}\rangle$, while the angle $\theta$ changes from $\theta_i$ to $\theta_f$. We next derive the relations connecting the various parameters of the pulse-sequence.

The total duration $T'$ (in the rescaled time $\tau$) is
\begin{equation}
\label{duration}
T'=2\tau_1+m\tau_2+(m-1)\tau_3,
\end{equation}
where $m=1,2,\ldots$ is the number of ``off" pulses in the sequence. Since the ``on" pulses have constant amplitude $u$, the total change in the angle is
\begin{equation*}
\theta_i-\theta_f=u[2\tau_1+(m-1)\tau_3],
\end{equation*}
thus
\begin{equation}
\label{first_relation}
2\tau_1+(m-1)\tau_3=\frac{\theta_i-\theta_f}{u}.
\end{equation}
Combining Eqs. (\ref{duration}), (\ref{first_relation}) we obtain
\begin{equation}
\label{second_relation}
\tau_2=\frac{1}{m}\left(T'-\frac{\theta_i-\theta_f}{u}\right).
\end{equation}

The next relation is derived from the requirement that the system should return to the adiabatic state $|\phi_{+}\rangle$ at the final time $\tau=T'$. Under the pulse-sequence $u(\tau)$, the propagator $U$ connecting the initial and final states, $\mathbf{b}(T')=U\mathbf{b}(0)$, can be expressed as
\begin{equation}
\label{total_propagator}
U=U_1W_2U_3\ldots W_2\,\mbox{or}\,U_3\ldots U_3W_2U_1,
\end{equation}
where $U_1, U_3$ are given by Eq. (\ref{U}), replacing there $T'$ with $\tau_1, \tau_3$, respectively, and
\begin{eqnarray}
\label{W}
W_2&=&e^{-i\frac{1}{2}\tau_2\sigma_z}\nonumber\\
&=& I\cos{\frac{\tau_2}{2}}-i\sin{\frac{\tau_2}{2}}\sigma_z.
\end{eqnarray}
The propagator in the middle of (\ref{total_propagator}) is $W_2$ or $U_3$, depending on the corresponding middle pulse.
Using the expressions for $U_1, W_2, U_3$ and the following property of Pauli matrices
\begin{equation}
\label{Pauli}
\sigma_a\sigma_b=\delta_{ab}I+i\epsilon_{abc}\sigma_c,
\end{equation}
where $a,b,c$ can be any of $x,y,z$, $\delta_{ab}$ is the Kronecker delta and $\epsilon_{abc}$ is the Levi-Civita symbol,
we can express the propagator $U$ as a linear combination of $\sigma_a$ and the identity $I$,
\begin{equation}
\label{propagator}
U=a_II+a_x\sigma_x+a_y\sigma_y+a_z\sigma_z.
\end{equation}
The coefficients of the matrices in the above expression are functions of the pulse-sequence parameters.

We next show that $a_x=0$. From Eqs. (\ref{propagator}), (\ref{total_propagator}), and a well-known identity regarding the trace of a matrix product, we have
\begin{eqnarray}
\label{a_x}
a_x&=&\frac{1}{2}\mbox{Tr}(\sigma_xU)\nonumber\\
&=&\frac{1}{2}\mbox{Tr}(\sigma_xU_1W_2U_3\ldots W_2\,\mbox{or}\,U_3\ldots U_3W_2U_1)\nonumber\\
&=&\frac{1}{2}\mbox{Tr}(\ldots U_3W_2U_1\sigma_xU_1W_2U_3\ldots W_2\,\mbox{or}\,U_3).
\end{eqnarray}
But, using the explicit expressions (\ref{U}), (\ref{W}) for $U_1, W_2, U_3$ and the identity (\ref{Pauli}), it is not hard to verify that
\begin{equation}
U_1\sigma_xU_1=W_2\sigma_xW_2=U_3\sigma_xU_3=\sigma_x.
\end{equation}
Using the above relations repeatedly in Eq. (\ref{a_x}), it is not difficult to see that the calculation of $a_x$ is reduced to the calculation of $\mbox{Tr}(\sigma_xW_2)$ or $\mbox{Tr}(\sigma_xU_3)$, depending whether the middle pulse is ``off" or ``on", respectively. But $\mbox{Tr}(\sigma_xW_2)=\mbox{Tr}(\sigma_xU_3)=0$, thus $a_x=0$ as well.

Now observe that $I, \sigma_z$ are diagonal. Since $a_x=0$ in Eq. (\ref{propagator}), if we set $a_y=0$ then $U$ is also diagonal. In this case, starting from $\mathbf{b}(0)=(1 \; 0)^T$ we find for the final state $\mathbf{b}(T')=U\mathbf{b}(0)$ that $b_2(T')=0$, and the system returns to the initial adiabatic state. The relation
\begin{equation}
\label{third_relation}
a_{y,m}(\tau_1,\tau_2,\tau_3,u)=0,
\end{equation}
along with Eqs. (\ref{first_relation}), (\ref{second_relation}), will be used for the determination of the pulse-sequence parameters. The subscript $m$ denotes that $a_y$ has a different functional form for different pulse-sequences, as we shall see in the examples of the next section.

Observe that Eqs. (\ref{first_relation}), (\ref{second_relation}) and (\ref{third_relation}) involve, aside the unknown durations $\tau_i$, $i=1,2,3$ and the amplitude $u$, the total duration $T'$, the initial and final angles $\theta_i, \theta_f$, and the number $m$ of ``off" pulses in the sequence. We clarify how these equations can be exploited to obtain the unknown characteristics of the pulse-sequence. First, the angles $\theta_i, \theta_f$ are given. Second, the total duration is also considered to be fixed and any value $T'>T'_0=\pi$ can be used, where the lower bound is obtained from Eq. (\ref{lower_bound}). Finally, number $m$ is determined as follows: for $T'_k<T'<T'_{k+1}$, where $T'_0=\pi$ and $T'_k$, $k=1,2,\ldots$ are the durations given in Eq. (\ref{Tk}), it is
\begin{equation}
\label{m}
m=k+1.
\end{equation}
The increasing number of pulses in the sequence as the total duration $T'$ passes through the values $T'_k$, $k=1,2,\ldots$, where the target adiabatic state is reached with a single constant pulse, is also inspired from minimum-time optimal control theory, see for example our previous work \cite{Stefanatos14}. Note also that the motivation for using longer durations is to find pulses with lower amplitude $u$, corresponding to less abrupt control signals. This goal is achieved when the number of the pulses in the sequence is determined as described above. For example, according to this rule, for durations $\pi<T'<T'_1$ we should use the simplest pulse sequence ``on-off-on".  If we keep the same pulse sequence for longer durations $T'>T'_1$ then, depending on the value of $T'$, the corresponding equation (\ref{third_relation}) may have no solutions or may have solutions for larger values of $u$ than the case where $\pi<T'<T'_1$.

We are left with four unknowns $\tau_i$, $i=1,2,3$ and $u$, and three equations, (\ref{first_relation}), (\ref{second_relation}), and (\ref{third_relation}). Using Eq. (\ref{first_relation}) we can express $\tau_3$ in terms of $\tau_1$ and $u$, while $\tau_2$ is Eq. (\ref{second_relation}) is already expressed as a function of $u$. If we plug these in Eq. (\ref{third_relation}), we obtain a (transcendental as we shall see) equation involving $\tau_1$ and $u$. The next step is the crucial one: we find the minimum value of the amplitude $u$ such that this transcendental equation has a positive solution for $\tau_1$. The physical motivation behind minimizing $u$ is to minimize the derivative $d\theta/d\tau$. From Eq. (\ref{second_relation}) observe that the requirement $\tau_2\geq 0$ implies that $u\geq(\theta_i-\theta_f)/T'$, thus the minimization of $u$ is a well defined problem.

We finally explain how to find the coefficient $a_y$ in Eq. (\ref{propagator}), which is equated to zero in Eq. (\ref{third_relation}). It is obtained from a relation similar to Eq. (\ref{a_x}),
\begin{eqnarray}
\label{a_y}
a_y&=&\frac{1}{2}\mbox{Tr}(\sigma_yU)\nonumber\\
&=&\frac{1}{2}\mbox{Tr}(\sigma_yU_1W_2U_3\ldots W_2\,\mbox{or}\,U_3\ldots U_3W_2U_1)\nonumber\\
&=&\frac{1}{2}\mbox{Tr}(\ldots U_3W_2U_1\sigma_yU_1W_2U_3\ldots W_2\,\mbox{or}\,U_3),
\end{eqnarray}
using repeatedly the equations
\begin{eqnarray}
U_1\sigma_yU_1&=&in_y\sin{\omega \tau_1}I+(n_z^2+n_y^2\cos{\omega \tau_1})\sigma_y\nonumber\\
&&+n_yn_z(1-\cos{\omega \tau_1})\sigma_z,\\
W_2\sigma_yW_2&=&\sigma_y,\\
W_2\sigma_zW_2&=&-i\sin{\tau_2}+\cos{\tau_2}\sigma_z,\\
U_3\sigma_yU_3&=&in_y\sin{\omega \tau_3}I+(n_z^2+n_y^2\cos{\omega \tau_3})\sigma_y\nonumber\\
&&+n_yn_z(1-\cos{\omega \tau_3})\sigma_z,\\
U_3\sigma_zU_3&=&-in_z\sin{\omega \tau_3}I+n_yn_z(1-\cos{\omega \tau_3})\sigma_y\nonumber\\
&&+(n_y^2+n_z^2\cos{\omega \tau_3})\sigma_z,
\end{eqnarray}
which can be derived from expressions (\ref{U}) for $U_1,U_3$ and (\ref{W}) for $W_2$, as well as property (\ref{Pauli}).

Having found the pulse-sequence $u(\tau)$ we can use Eq. (\ref{control}) to find $\theta(\tau)$ and then Eq. (\ref{rescaled_time}) to find $\theta(t)$ in the original time $t$. In the following section we provide examples which elucidate the procedure described above for obtaining the pulse-sequence when $\theta_i,\theta_f$ and $T'$ are given.

\section{Examples}

\label{sec:examples}

In all the following examples we consider a change in the detuning $\Delta$ from $-10\Omega$ to $10\Omega$, as in \cite{Martinis14}, corresponding to
\begin{equation}
\label{angle_values}
\theta_i=\pi-\theta_f,\quad\theta_f=\tan^{-1}\frac{1}{10}.
\end{equation}
For these values of the initial and final angles we obtain from Eq. (\ref{Tk})
\begin{equation}
\label{Tk_values}
T'_1=1.77\pi,\quad T'_2=3.89\pi,\quad T'_3=5.93\pi.
\end{equation}

\subsection{On-Off-On}

\begin{figure}[t]
 \centering
		\begin{tabular}{c}
     	
      \subfigure[$\ $]{
	            \label{fig:fidelity1}
	            \includegraphics[width=.7\linewidth]{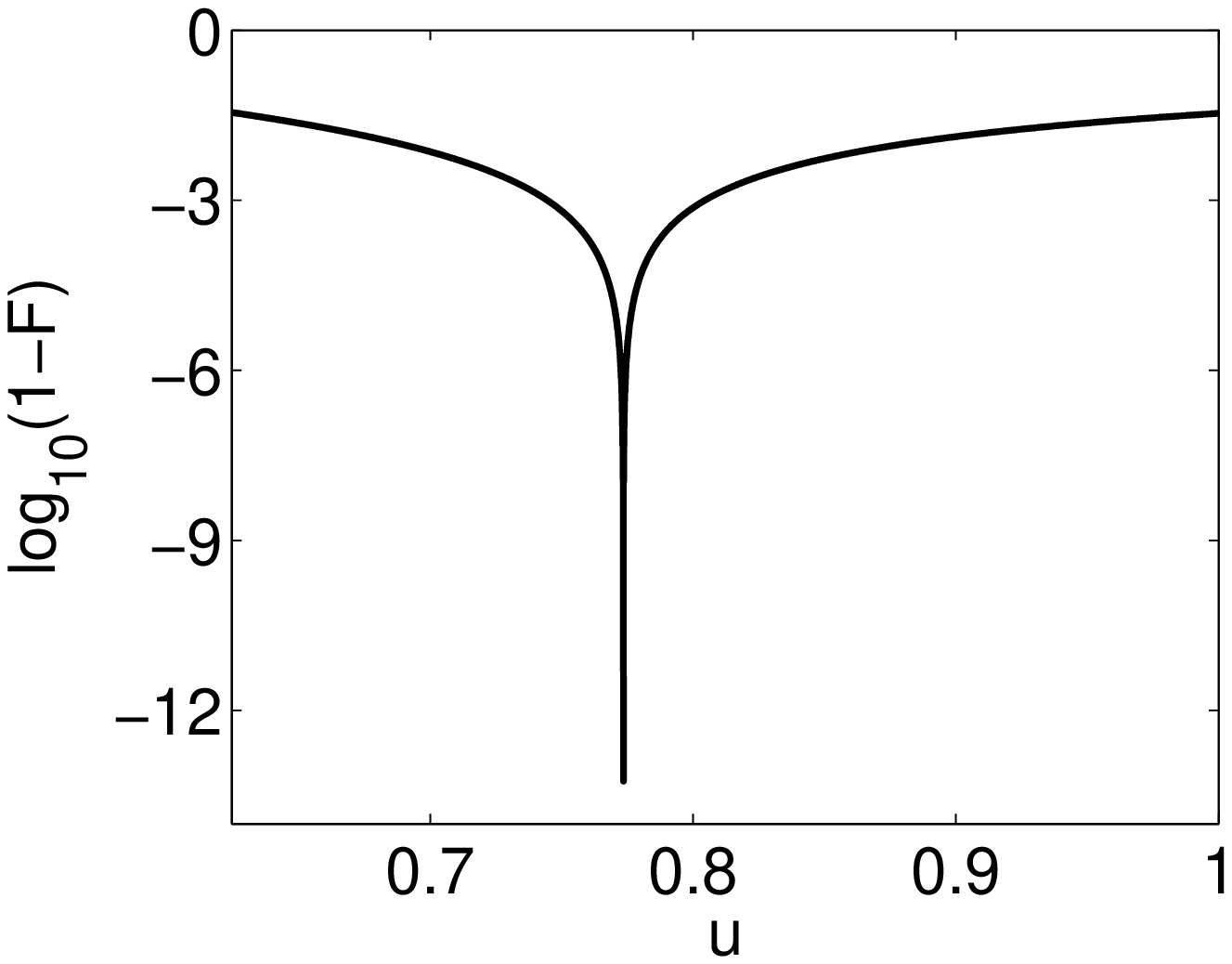}} \\
      \subfigure[$\ $]{
	            \label{fig:fidelity2}
	            \includegraphics[width=.7\linewidth]{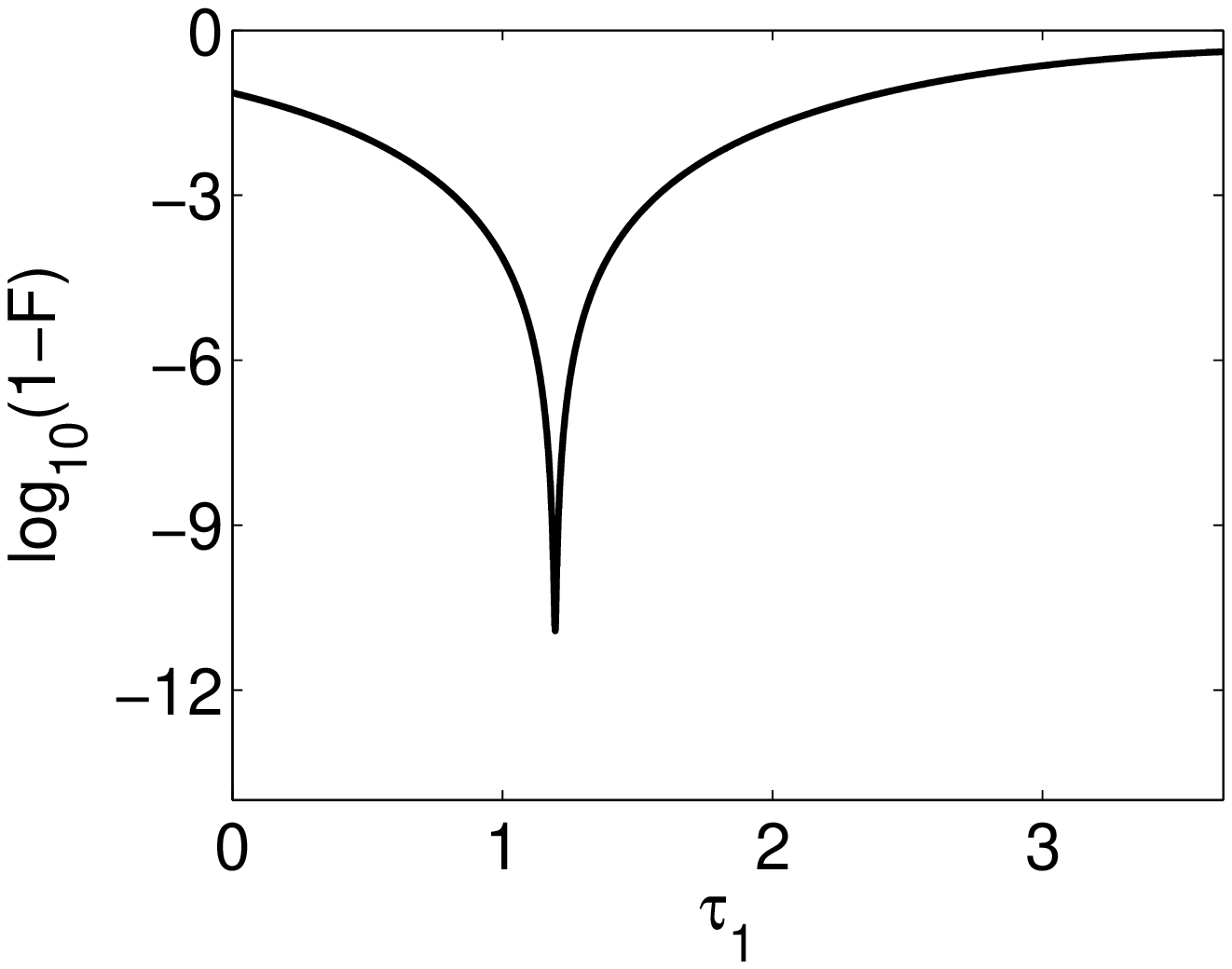}} \\
       \subfigure[$\ $]{
	            \label{fig:fidelity3}
	            \includegraphics[width=.7\linewidth]{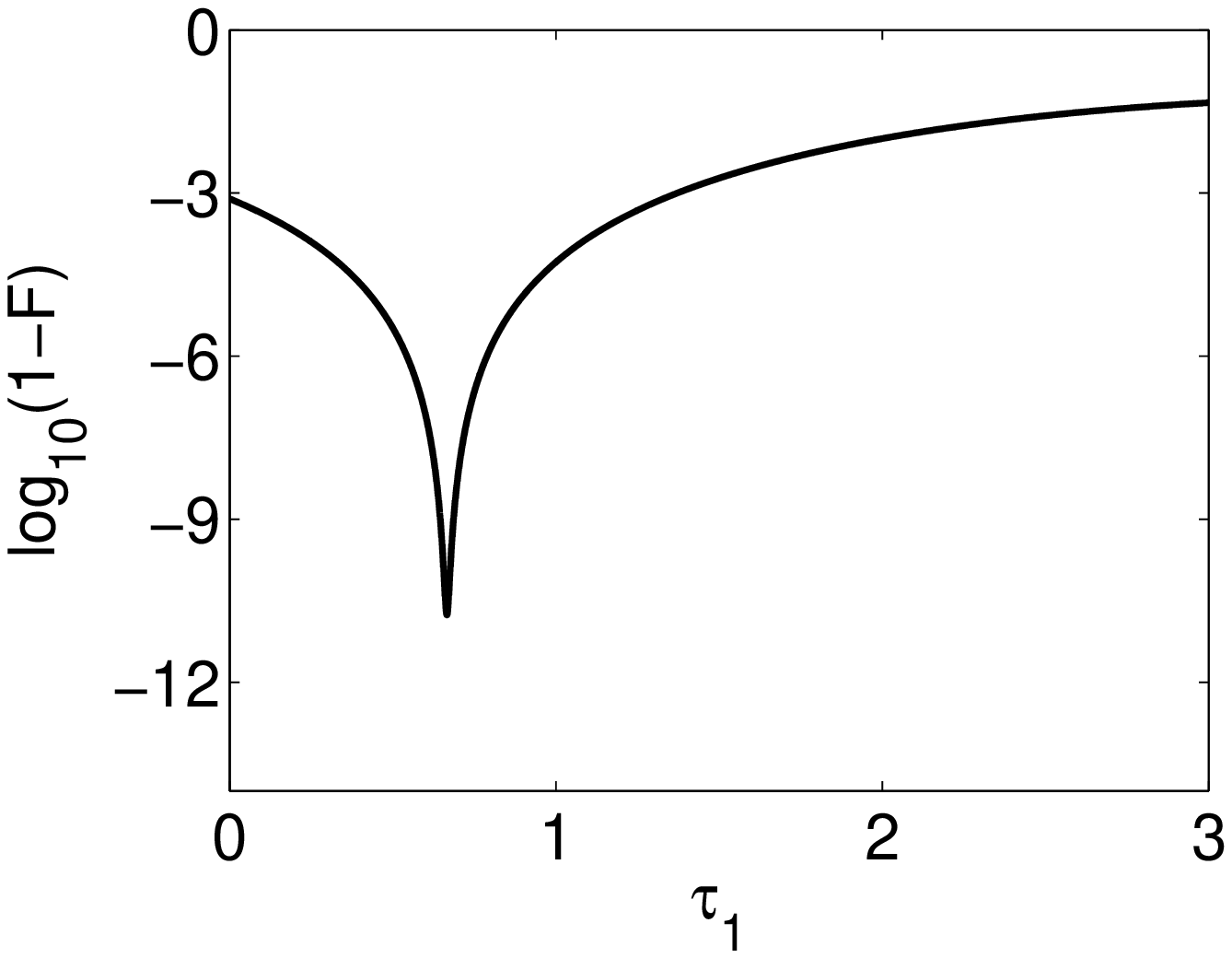}}
		\end{tabular}
\caption{(Color online) Logarithmic error of the final state (a) for the ``on-off-on" pulse-sequence with total duration $T'=1.5\pi$ in rescaled time, as a function of the pulse amplitude $u$,  (b) the ``on-off-on-off-on" pulse-sequence with total duration $T'=3\pi$ in rescaled time, as a function of the duration $\tau_1$ of the first pulse, (c) the ``on-off-on-off-on-off-on" pulse-sequence with total duration $T'=4.5\pi$ in rescaled time, as a function of the duration $\tau_1$ of the first pulse. The resonances are clearly displayed.}
\label{fig:resonances}
\end{figure}

We first find the pulse sequence with total duration $T'=1.5\pi$ in the rescaled time $\tau$. Since $T'_0<T'<T'_1$, the pulse sequence contains $m=1$ ``off" pulses, thus it has the form ``on-off-on". Only for this special form, where there are no intermediate ``on" pulses thus $\tau_3=0$, we have only three unknowns, $\tau_1, \tau_2, u$, in the three equations (\ref{first_relation}), (\ref{second_relation}), (\ref{third_relation}). In this case we can simply solve the equations; the minimization procedure with respect to $u$ described in the previous section in not necessary.  Eq. (\ref{first_relation}) gives
\begin{equation}
\label{t1_u}
\tau_1=\frac{\theta_i-\theta_f}{2u},
\end{equation}
thus both $\tau_1, \tau_2$ are expressed as functions of $u$. On the other hand, Eq. (\ref{a_y}) becomes
\begin{eqnarray}
\label{ay1}
a_{y,1}&=&\frac{1}{2}\mbox{Tr}(\sigma_yU)=\frac{1}{2}\mbox{Tr}(\sigma_yU_1W_2U_1)=\frac{1}{2}\mbox{Tr}(U_1\sigma_yU_1W_2)\nonumber\\
&=&2i n_y\sin{(\omega \tau_1/2)}\times\nonumber\\
&&\big[\cos{(\omega \tau_1/2)}\cos{(\tau_2/2)}-n_z\sin{(\omega \tau_1/2)}\sin{(\tau_2/2)}\big].\nonumber\\
\end{eqnarray}
We can perform a quick consistency check of the above expression by setting $\tau_2=0$, in which case the ``on-off-on" pulse sequence degenerates to a constant ``on" pulse of duration $T'=2\tau_1$. We obtain $a_{y,1}=i n_y\sin{\omega \tau_1}$, which is indeed the corresponding coefficient for a constant pulse of duration $T'=2\tau_1$, see Eq. (\ref{U}). If we plug in Eq. (\ref{ay1}) the expressions (\ref{t1_u}), (\ref{second_relation}) for $\tau_1, \tau_2$, we find $a_{y,1}$ as a function of parameter $u$ only. In Fig. \ref{fig:fidelity1} we plot the logarithmic error
\begin{equation}
\label{logarithmic_error_1}
\log_{10}{(1-F)}=\log_{10}{|b_2(T')|^2}=\log_{10}{|a_{y,1}|^2}
\end{equation}
as a function of $u$, for the total duration $T'=1.5\pi$ that we use here. The resonance is observed at the value $u=0.773436$, which is also the solution of the transcendental equation $a_{y,1}=0$.

\begin{figure}[t]
 \centering
		\begin{tabular}{c}
     	
      \subfigure[$\ $]{
	            \label{fig:fidelity4}
	            \includegraphics[width=.7\linewidth]{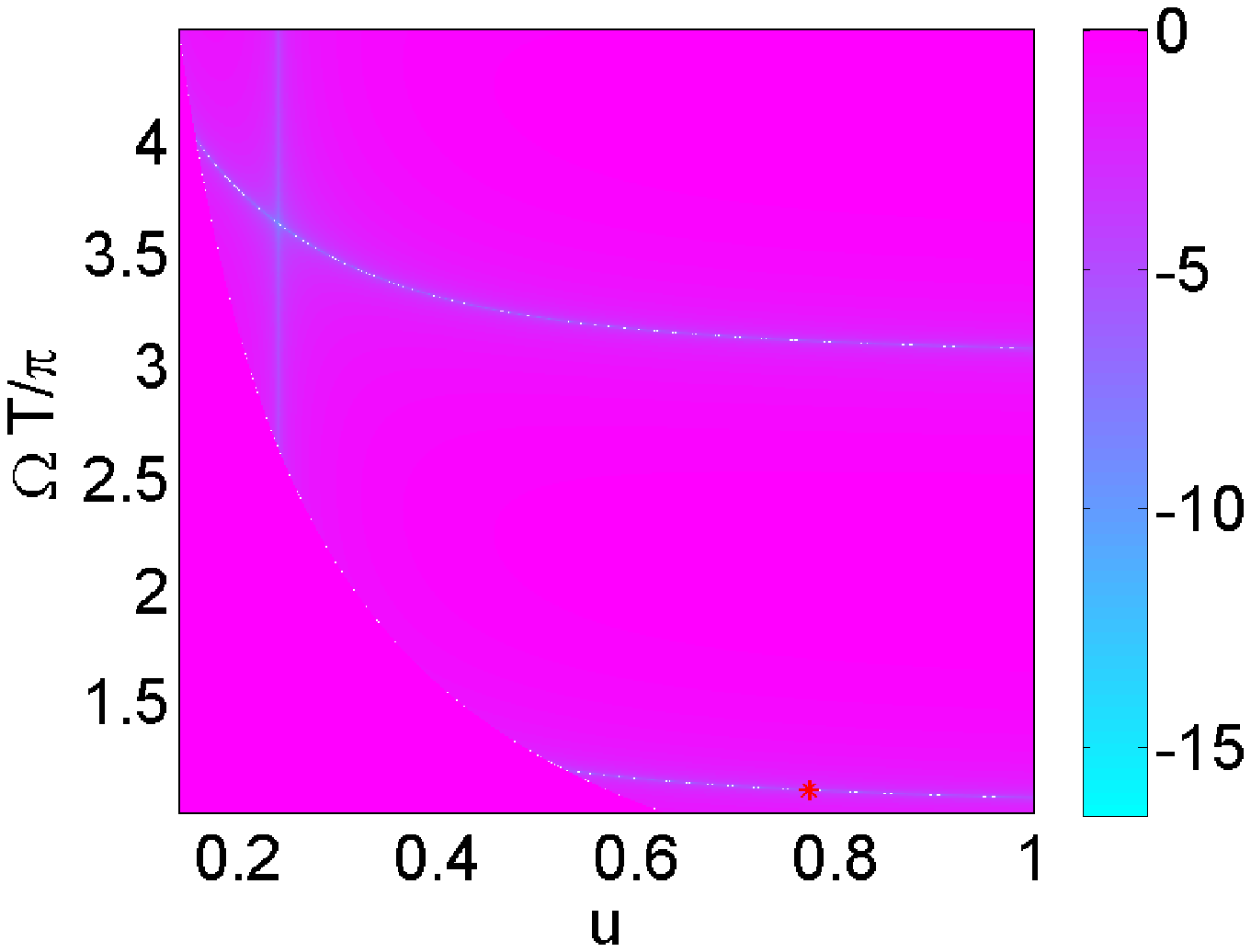}} \\
       \subfigure[$\ $]{
	            \label{fig:fidelity5}
	            \includegraphics[width=.7\linewidth]{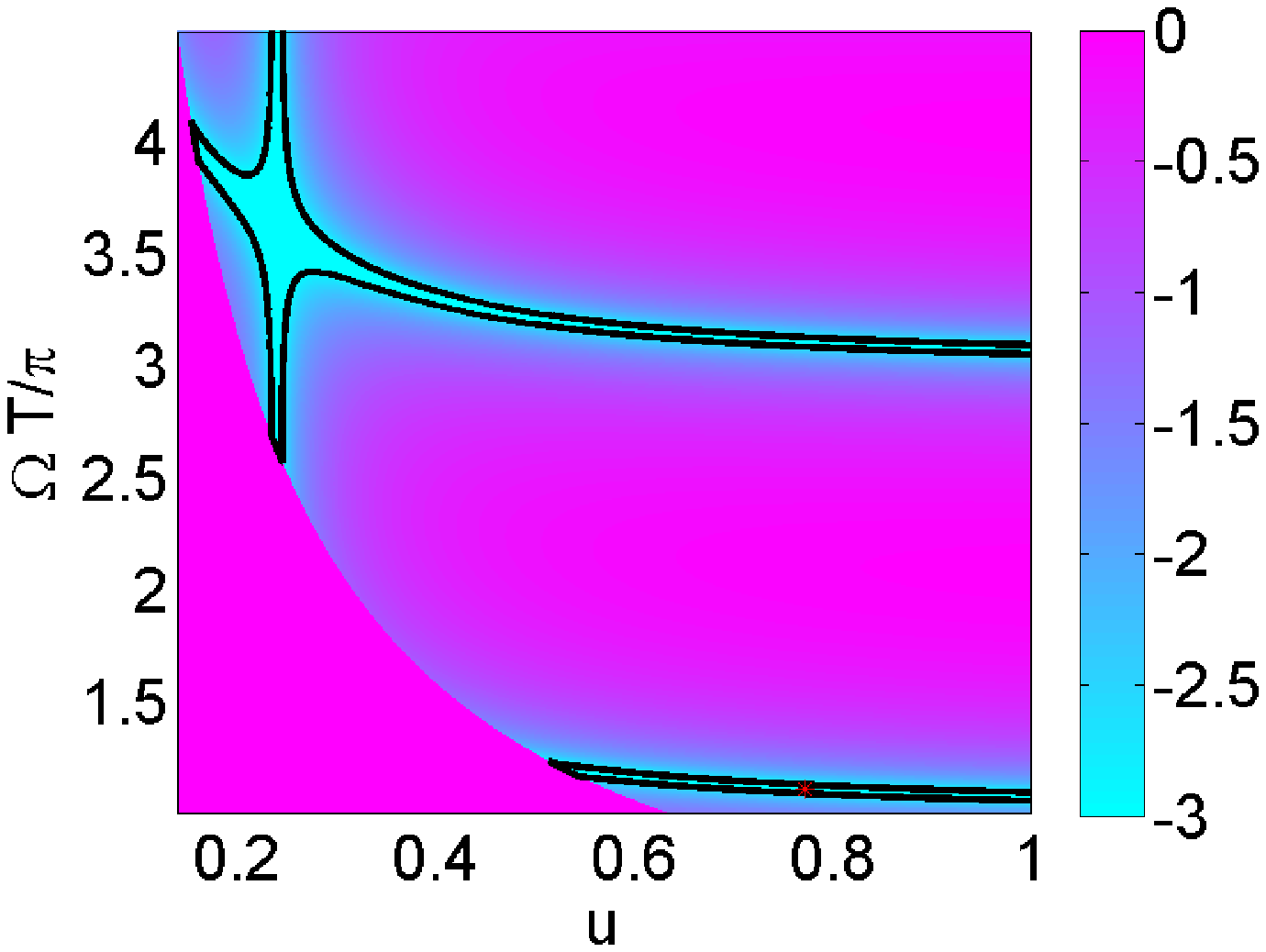}}
		\end{tabular}
\caption{(Color online) (a) Logarithmic error of the final state for the ``on-off-on" pulse-sequence as a function of the pulse amplitude $u$ and the total duration $T$ in original time. The red star in the lower horizontal branch corresponds to the resonance in Fig. \ref{fig:fidelity1}. (b) Same as in (a) but with a different color scale. Here, all the points satisfying $\log_{10}{(1-F)}\leq-3$ are displayed with cyan color, at the bottom of the scale, while the black solid line denotes the contour of this region.}
\label{fig:fidelity}
\end{figure}

The durations $\tau_1, \tau_2$ can be obtained from Eq. (\ref{t1_u}), (\ref{second_relation}), respectively.
We explain how to obtain the corresponding durations $t_1, t_2$ in the original time $t$. Observe that during the initial ``on" pulse the evolution of angle $\theta$ in the rescaled and original times is given by Eqs. (\ref{angle_rescaled}) and (\ref{angle_original}), respectively, where of course $u_k$ is replaced by $u$. At the switching time $\tau=\tau_1$ it is
\begin{equation*}
\theta(\tau_1)=\theta_i-u\tau_1=\frac{\theta_i+\theta_f}{2}=\frac{\pi}{2},
\end{equation*}
where we have used Eq. (\ref{t1_u}) and the symmetry $\theta_i=\pi-\theta_f$ between the initial and final angles. In the original time $t$ we have
\begin{equation*}
\theta(t_1)=\cos^{-1}(\cos{\theta_i}+u\Omega t_1).
\end{equation*}
But $\theta(t_1)=\theta(\tau_1)=\pi/2$, thus
\begin{equation}
\label{t1}
t_1=-\frac{\cos{\theta_i}}{u}\cdot\frac{1}{\Omega}.
\end{equation}
During the ``off" pulse the angle maintains the constant value $\pi/2$. From Eqs. (\ref{rescaled_time}) and (\ref{second_relation}) we easily obtain
\begin{equation}
\label{t2}
t_2=\frac{\tau_2}{\Omega}=\left(T'-\frac{\theta_i-\theta_f}{u}\right)\cdot\frac{1}{\Omega}.
\end{equation}
Note that Eqs. (\ref{t1}), (\ref{t2}) hold for the symmetric case $\theta_i=\pi-\theta_f$. If $\theta_i, \theta_f$ are not symmetric, then similar equations can be easily obtained.

We can exploit these equations and plot the logarithmic error (\ref{logarithmic_error_1}) as a function of both $u$ and the total duration $T=2t_1+t_2$ (in the original time $t$). Such a plot is displayed in Fig. \ref{fig:fidelity4}. The cyan lines correspond to the solutions of $a_{y,1}=0$, while the boundary hyperbola in the $u-T$ plane is defined by $T\geq2t_1=-2\cos{\theta_i}/(\Omega u)$. The vertical cyan line corresponds to the nullification of the first factor of $a_{y,1}$, see Eq. (\ref{ay1}), while the horizontal lines correspond to the nullification of the second factor. The lower horizontal line represents minimum-time solutions corresponding to the first resonance of this second factor. The solution highlighted with red star corresponds to the resonance shown in Fig. \ref{fig:fidelity1}, with duration $T'=1.5\pi$ in the rescaled time $\tau$ and $T=2t_1+t_2\approx 1.108\pi/\Omega$ in the original time $t$. The solutions on the other horizontal branch have longer durations $T>T_1$ for the same amplitude $u$, thus they are not time-optimal, yet we display them for completeness. They correspond to the second resonance of the second factor of $a_{y,1}$.

The solutions on the vertical line are also not time-optimal. They correspond to the nullification of the first factor of $a_{y,1}$, thus they satisfy $\sin{(\omega \tau_1/2)}=0$, where $\tau_1$ is given in Eq. (\ref{t1_u}). For the values of $\theta_i, \theta_f$ that we use here it turns out that $u\approx 0.2408$, the value where the vertical line lies in the figure. Note that the durations of these solutions satisfy $T\geq T_2\approx 2.63\pi/\Omega$, since the fastest solution in the family is simply the second resonance with constant amplitude, given by Eqs. (\ref{uk}) and (\ref{durations_original}) for $k=2$. The other solutions in the family have the same $t_1$ and increasing duration $t_2$ of the ``off" pulse, thus an increasing total duration. It is not hard to actually visualize these solutions on the Bloch sphere, especially for symmetric $\theta_i, \theta_f$. In the original reference frame, the first ``on" pulse brings the total field to $\theta=\pi/2$ and aligns the state vector with the total field. During the subsequent ``off" pulse the vectors remain aligned in this position for duration $t_2$. Under the final ``on" pulse both vectors evolve symmetrically to the previous case and align at the final angle $\theta_f$. In the adiabatic frame, where the total field points constantly to the north pole, under the first ``on" pulse the state vector performs a full rotation and returns to the north pole. It remains there for the duration of the ``off" pulse. The final ``on" pulse drives a second full rotation of the state vector, before it returns to the north pole.

Observe in Fig. \ref{fig:fidelity4} that, as the duration increases, the vertical line intersects the upper horizontal line. At this point, both factors of $a_{y,1}$ become zero. The corresponding amplitude value is that of the vertical line, $u\approx 0.2408$. The corresponding duration is obtained from the nullification of the second factor in Eq. (\ref{ay1}). Since $\sin{(\omega \tau_1/2)}=0$ (the first factor is also zero), the second factor is proportional to $\cos{(\tau_2/2)}$ and becomes zero for $\tau_2=\pi$. But during the ``off" pulse and for symmetric initial and final angles it is $\theta=\pi/2$, thus the corresponding duration in the original time is $t_2=\pi/\Omega$. The total duration at the intersection point is $T=T_2+\pi/\Omega\approx 3.63\pi/\Omega$. Obviously, the robustness of the transfer is increased around this point. This becomes apparent in Fig. \ref{fig:fidelity5}, which is similar to Fig. \ref{fig:fidelity4} but with a different color scale. Here, all the points satisfying $\log_{10}{(1-F)}\leq-3$ are displayed with cyan color, at the bottom of the scale, while the black solid line denotes the contour of this region.

\begin{figure*}[t]
 \centering
		\begin{tabular}{cc}
     	\subfigure[$\ $]{
	            \label{fig:pulse1}
	            \includegraphics[width=.45\linewidth]{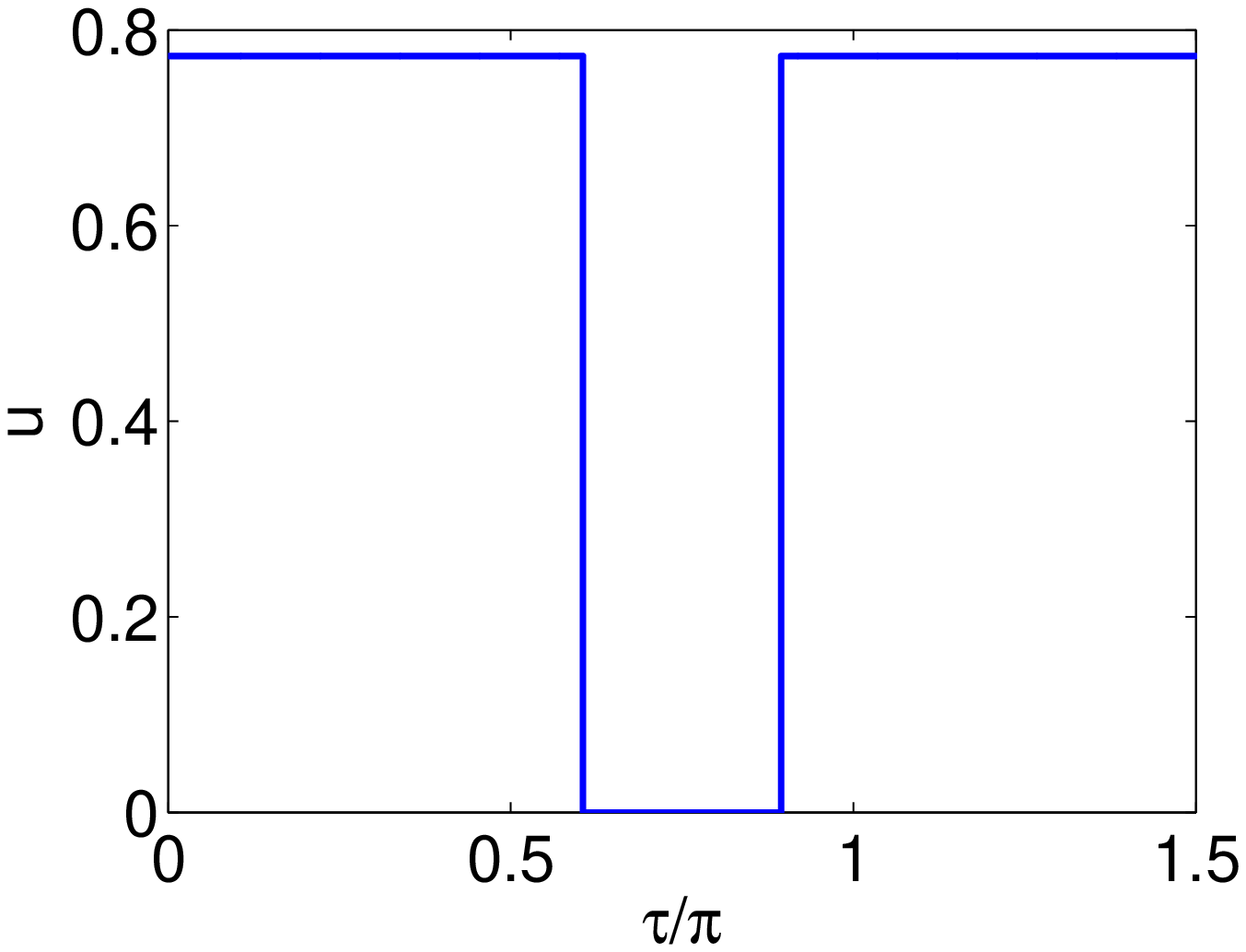}} &
       \subfigure[$\ $]{
	            \label{fig:theta1}
	            \includegraphics[width=.45\linewidth]{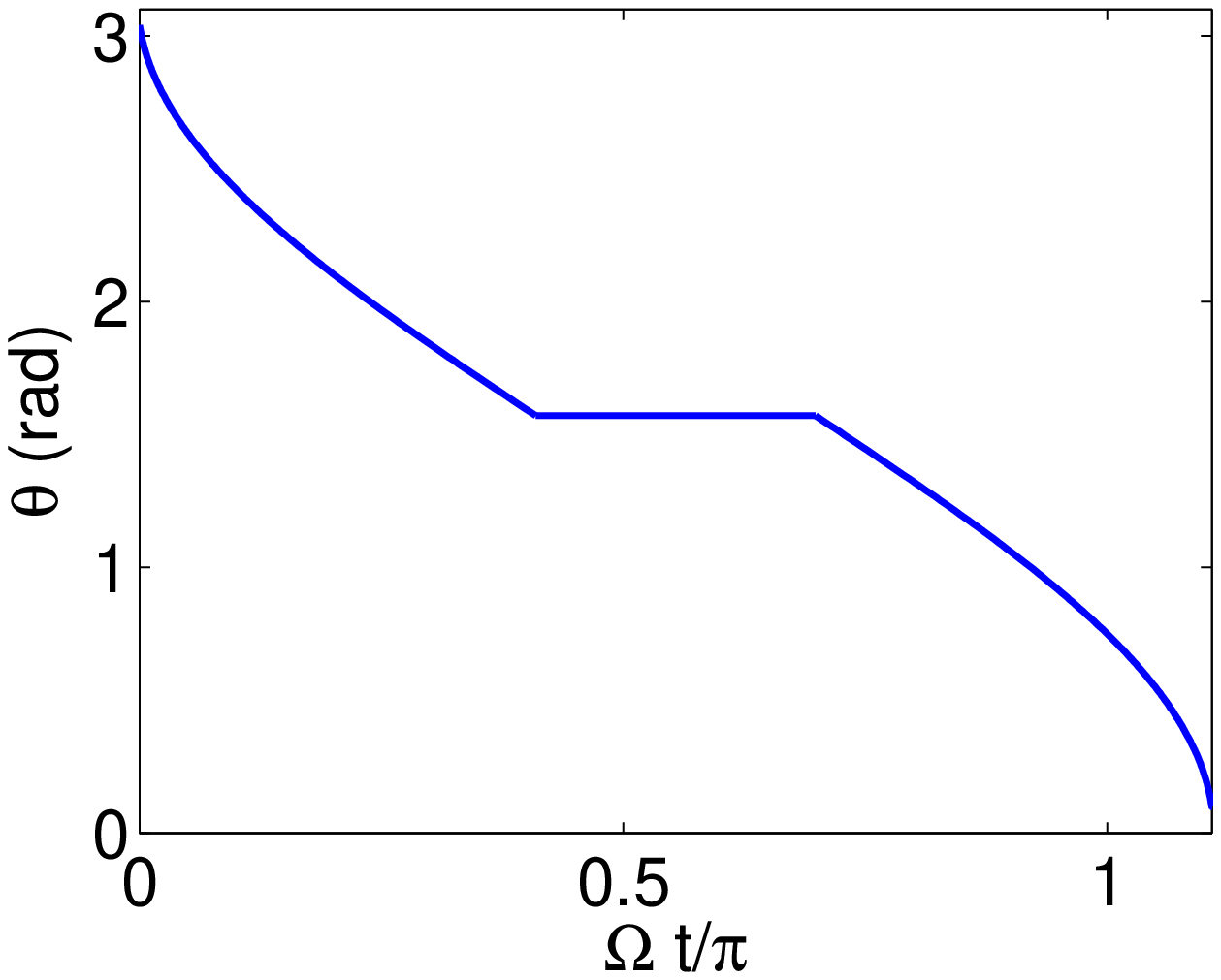}} \\
      \subfigure[$\ $]{
	            \label{fig:original1}
	            \includegraphics[width=.45\linewidth]{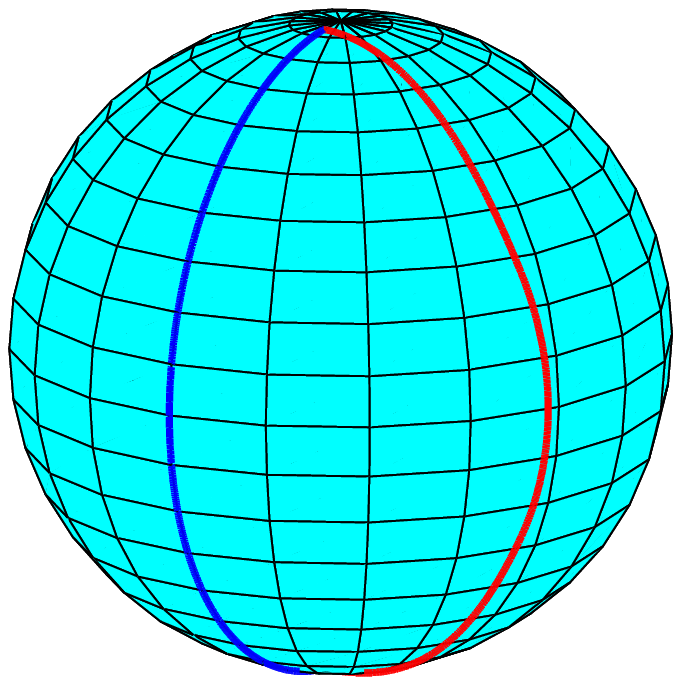}} &
       \subfigure[$\ $]{
	            \label{fig:adiabatic1}
	            \includegraphics[width=.45\linewidth]{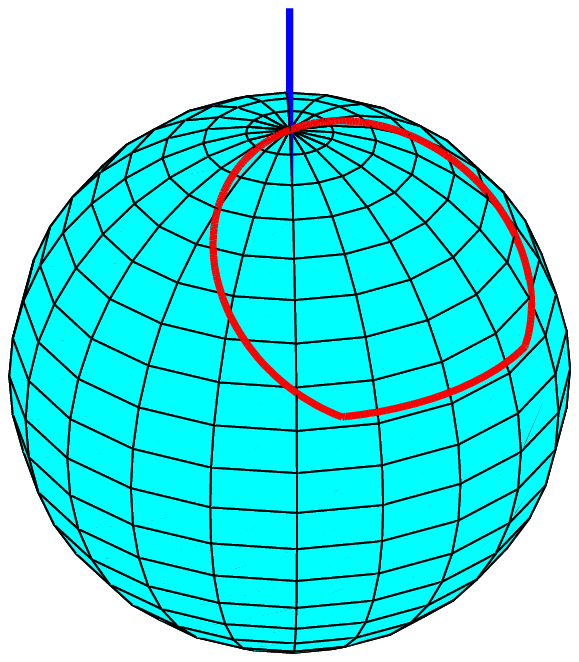}}
		\end{tabular}
\caption{(Color online) (a) Pulse sequence $u(\tau)$ in the rescaled time $\tau$, corresponding to duration $T'=1.5\pi$. (b) Corresponding evolution of the total field angle $\theta(t)$ in the original time $t$, with duration $T=2t_1+t_2\approx 1.108\pi/\Omega$. (c) State trajectory (red solid line) on the Bloch sphere in the original reference frame. The blue solid line on the meridian lying on the $xz$-plane indicates the change in the total field angle $\theta$. (d) State trajectory (red solid line) on the Bloch sphere in the adiabatic frame. Observe that in this frame the state of the system returns to the north pole, while the total field points constantly in the $\hat{z}$-direction (blue solid line).}
\label{fig:one_off}
\end{figure*}

We now return to the minimum-time solution displayed as a resonance in Fig. \ref{fig:fidelity1} and highlighted with a red star in the lower parallel branch of Fig. \ref{fig:fidelity}. In Fig. \ref{fig:pulse1} we plot the pulse sequence $u(\tau)$ in the rescaled time $\tau$, where recall that the corresponding duration is $T'=1.5\pi$. In Fig. \ref{fig:theta1} we show the corresponding evolution of the total field angle $\theta(t)$ in the original time $t$, where the duration corresponding to $T'$ is $T=2t_1+t_2\approx 1.108\pi/\Omega$. The detuning $\Delta(t)$ in the original time $t$ is displayed in Fig. \ref{fig:detuning1}. In Fig. \ref{fig:original1} we plot with red solid line the corresponding state trajectory on the Bloch sphere and in the original reference frame. The blue solid line on the meridian indicates the change in the total field angle $\theta$. Finally, in Fig. \ref{fig:adiabatic1} we plot the same trajectory (red solid line) but in the adiabatic frame. Note that in this frame the system starts from the adiabatic state at the north pole and returns there at the final time, while the total field points constantly in the $\hat{z}$-direction (blue solid line).

\subsection{On-Off-On-Off-On}

\begin{figure*}[t]
 \centering
		\begin{tabular}{cc}
     	\subfigure[$\ $]{
	            \label{fig:pulse2}
	            \includegraphics[width=.45\linewidth]{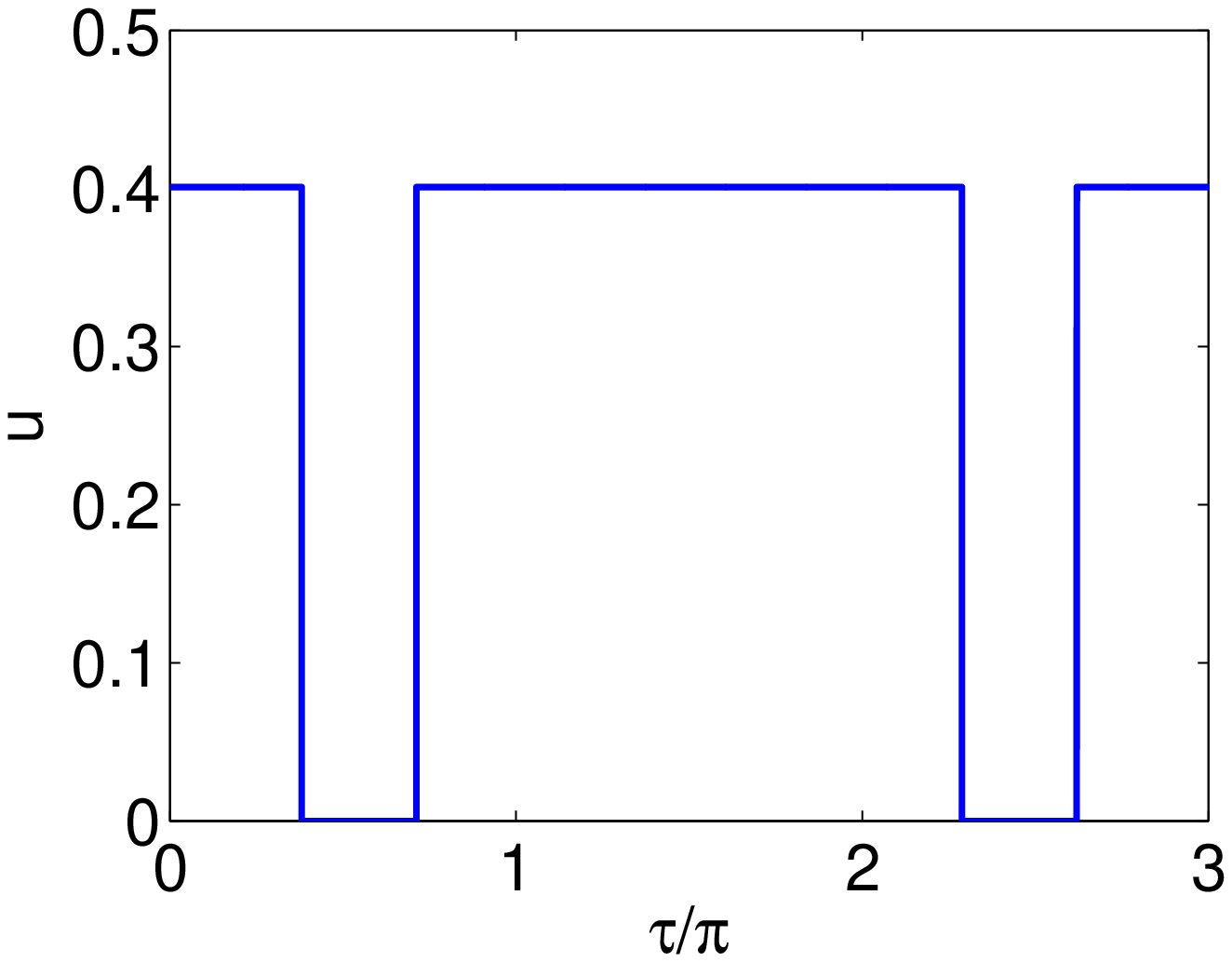}} &
       \subfigure[$\ $]{
	            \label{fig:theta2}
	            \includegraphics[width=.45\linewidth]{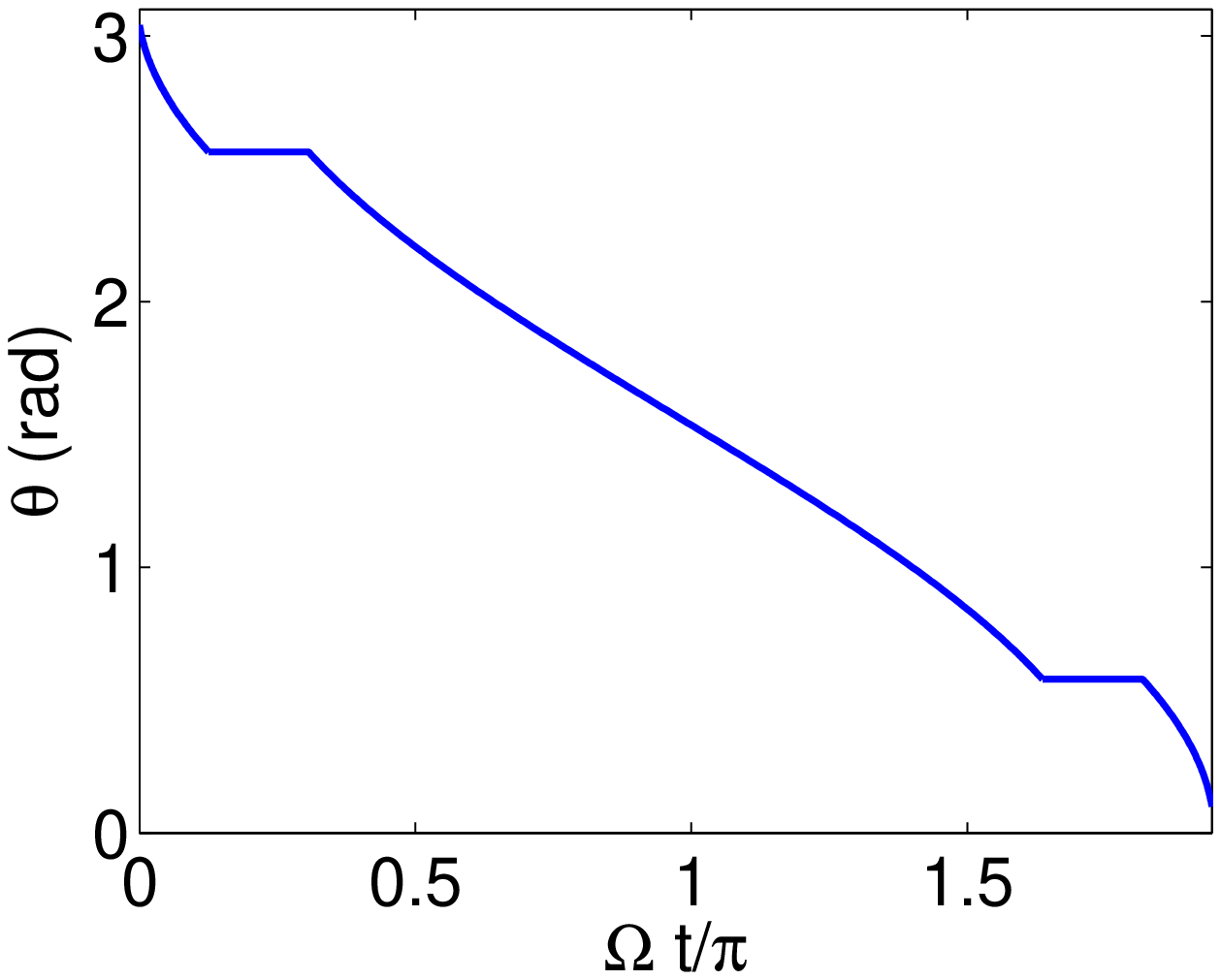}} \\
      \subfigure[$\ $]{
	            \label{fig:original2}
	            \includegraphics[width=.45\linewidth]{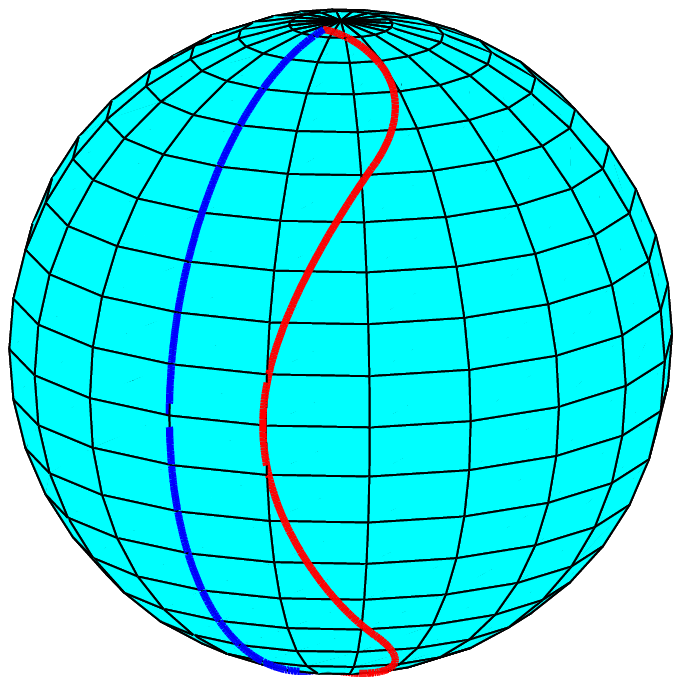}} &
       \subfigure[$\ $]{
	            \label{fig:adiabatic2}
	            \includegraphics[width=.45\linewidth]{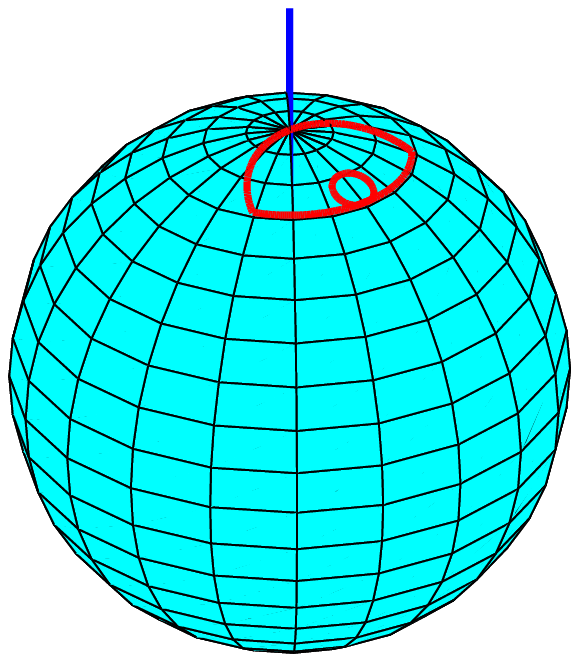}}
		\end{tabular}
\caption{(Color online) (a) Pulse sequence $u(\tau)$ in the rescaled time $\tau$, corresponding to duration $T'=3\pi$. (b) Corresponding evolution of the total field angle $\theta(t)$ in the original time $t$, with duration $T\approx 1.943\pi/\Omega$. (c) State trajectory (red solid line) on the Bloch sphere in the original reference frame. The blue solid line on the meridian lying on the $xz$-plane indicates the change in the total field angle $\theta$. (d) State trajectory (red solid line) on the Bloch sphere in the adiabatic frame.}
\label{fig:two_off}
\end{figure*}

We find next the pulse-sequence with total duration $T'=3\pi$ in the rescaled time $\tau$. Since $T'_1<T'<T'_2$, the pulse sequence contains $m=2$ ``off" pulses, thus it has the form ``on-off-on-off-on", see Fig. \ref{fig:pulse2}. We follow the procedure described in Section \ref{sec:generalized}. Eq. (\ref{a_y}) becomes
\begin{widetext}
\begin{eqnarray}
\label{ay2}
a_{y,2}&=&\frac{1}{2}\mbox{Tr}(\sigma_yU)=\frac{1}{2}\mbox{Tr}(\sigma_yU_1W_2U_3W_2U_1)=\frac{1}{2}\mbox{Tr}(W_2U_1\sigma_yU_1W_2U_3)\nonumber\\
&=&in_y\cos{(\omega \tau_3/2)}\big[\sin{\omega \tau_1}\cos{\tau_2}-n_z\sin{\tau_2}(1-\cos{\omega \tau_1})\big]+\nonumber\\
&&in_y\sin{(\omega \tau_3/2)}\bigg\{\cos{\omega \tau_1}+n_z\big[-\sin{\omega \tau_1}\sin{\tau_2}+n_z(1-\cos{\omega \tau_1})(1-\cos{\tau_2})\big]\bigg\}.\nonumber\\
\end{eqnarray}
\end{widetext}
We can quickly check this complicated expression, as we did for Eq. (\ref{ay1}) in the previous example. If we set $\tau_2=0$, the ``on-off-on-off-on" pulse-sequence degenerates to a constant ``on" pulse of duration $T'=2\tau_1+\tau_3$. Eq. (\ref{ay2}) reduces to $a_{y,2}=in_y\sin{\omega (\tau_1+\tau_3/2)}$, which is indeed the right expression for a constant pulse of duration $T'=2\tau_1+\tau_3$, see Eq. (\ref{U}). Another consistency check can be performed by setting $\tau_3=0$, in which case the pulse-sequence degenerates to a ``on-off-on" sequence where the ``off" pulse has duration $2\tau_2$. It is not hard to verify that Eq. (\ref{ay2}) reduces to the form (\ref{ay1}) with $2\tau_2$ in place of $\tau_2$.

If we use Eqs. (\ref{first_relation}), (\ref{second_relation}) for $\tau_3, \tau_2$, respectively, and plug the corresponding expressions in Eq. (\ref{ay2}), we obtain $a_{y,2}$ as a function of duration $\tau_1$ and the amplitude $u$ of the pulse-sequence. For the total duration $T'=3\pi$ that we use here, the minimum value of $u$ for which the transcendental equation $a_{y,2}=0$ has a solution with respect to $\tau_1$ is found numerically to be $u=0.40089$. In Fig. \ref{fig:fidelity2} we plot the logarithmic error
\begin{equation}
\log_{10}{(1-F)}=\log_{10}{|b_2(T')|^2}=\log_{10}{|a_{y,2}|^2}
\end{equation}
as a function of $\tau_1$, for $u=0.40089$ and $T'=3\pi$. The observed resonance corresponds to the solution of the transcendental equation $a_{y,2}=0$.

In Fig. \ref{fig:pulse2} we plot the pulse sequence $u(\tau)$ corresponding to duration $T'=3\pi$ in the rescaled time $\tau$. In Fig. \ref{fig:theta2} we show the corresponding evolution of the total field angle $\theta(t)$ in the original time $t$, where the duration corresponding to $T'$ is found by integrating Eq. (\ref{rescaled_time}) to be $T\approx 1.943\pi/\Omega$. The detuning $\Delta(t)$ in the original time $t$ is displayed in Fig. \ref{fig:detuning2}. In Fig. \ref{fig:original2} we plot with red solid line the corresponding state trajectory on the Bloch sphere and in the original reference frame. The blue solid line on the meridian indicates the change in the total field angle $\theta$. Finally, in Fig. \ref{fig:adiabatic2} we plot the same trajectory (red solid line) but in the adiabatic frame, where the total field points constantly in the $\hat{z}$-direction (blue solid line). Observe that the trajectory stays closer to the north pole than in the previous case. The reason is that the $y$-component of the effective field, which is $-u$, see Eqs. (\ref{rescaled_adiabatic_H}) and (\ref{control}), is now smaller than before. Also note that the trajectory in this frame contains a loop, which might look surprising at first sight, especially if you think that this problem is actually connected to minimum-time optimal control, as mentioned in the previous section. The catch here is that there is actually an extra state variable not shown in this frame, the angle $\theta$, which evolves from $\theta_i$ to $\theta_f$. If the trajectory is displayed in the higher-dimensional space of all the state variables, the loop disappears.

\subsection{On-Off-On-Off-On-Off-On}

\begin{figure*}[t]
 \centering
		\begin{tabular}{cc}
     	\subfigure[$\ $]{
	            \label{fig:pulse3}
	            \includegraphics[width=.45\linewidth]{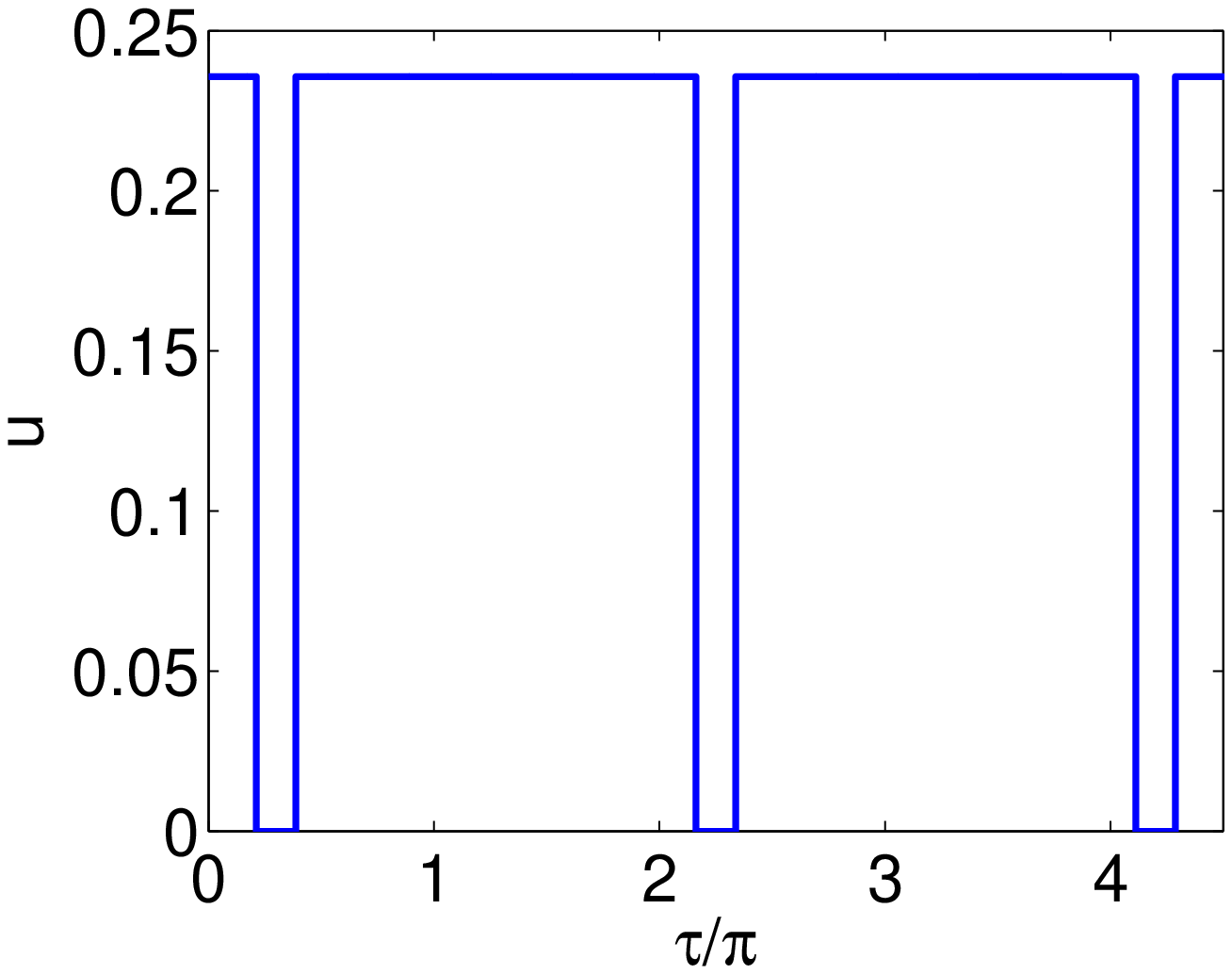}} &
       \subfigure[$\ $]{
	            \label{fig:theta3}
	            \includegraphics[width=.45\linewidth]{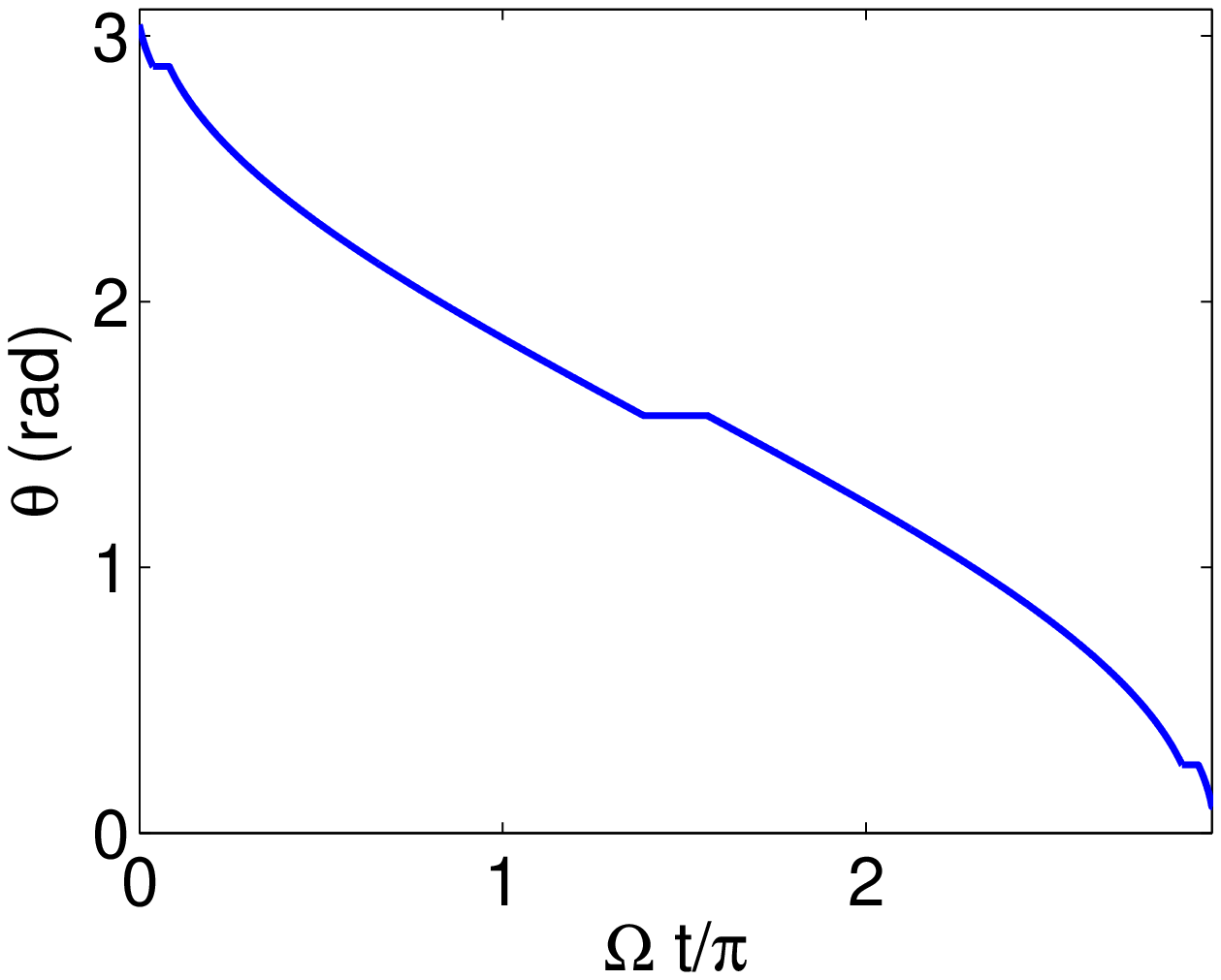}} \\
      \subfigure[$\ $]{
	            \label{fig:original3}
	            \includegraphics[width=.45\linewidth]{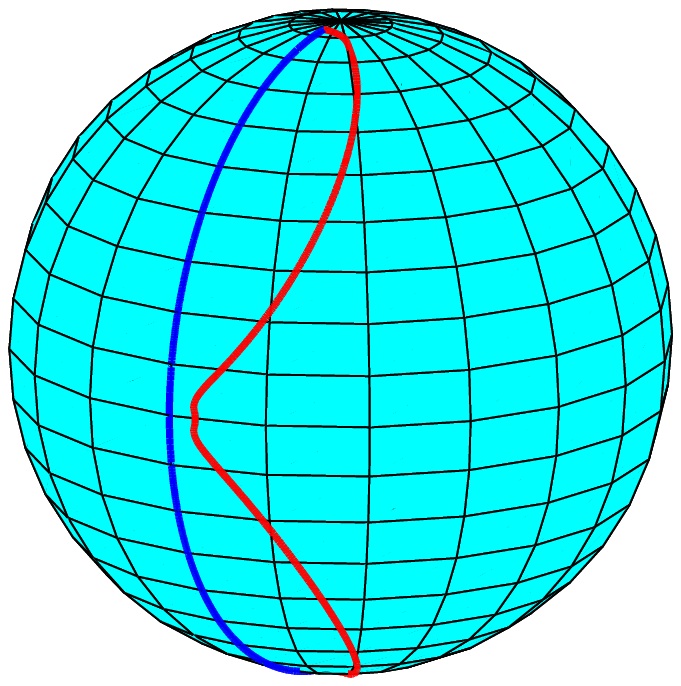}} &
       \subfigure[$\ $]{
	            \label{fig:adiabatic3}
	            \includegraphics[width=.45\linewidth]{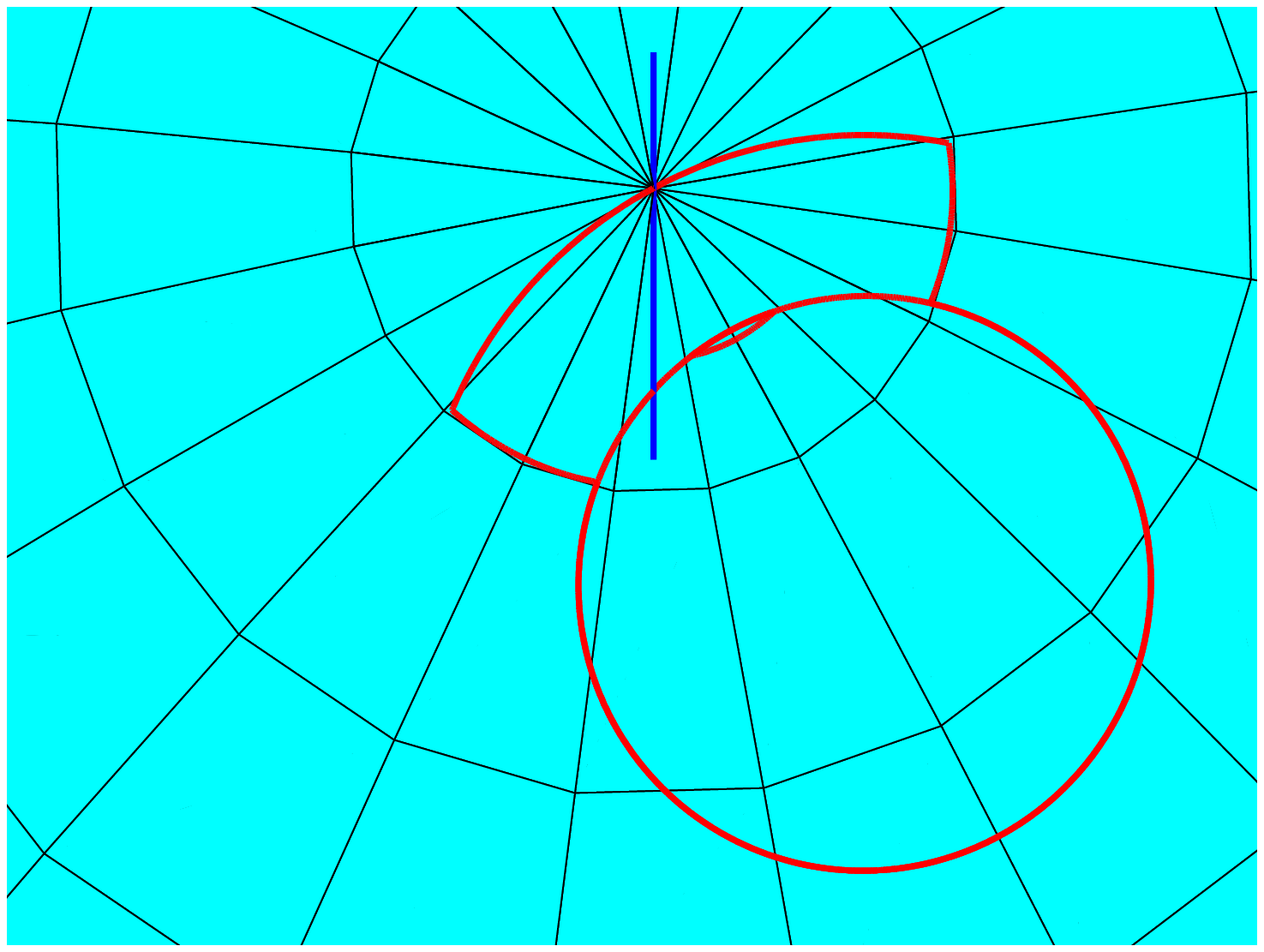}}
		\end{tabular}
\caption{(Color online) (a) Pulse sequence $u(\tau)$ in the rescaled time $\tau$, corresponding to duration $T'=4.5\pi$. (b) Corresponding evolution of the total field angle $\theta(t)$ in the original time $t$, with duration $T\approx 2.952\pi/\Omega$. (c) State trajectory (red solid line) on the Bloch sphere in the original reference frame. The blue solid line on the meridian lying on the $xz$-plane indicates the change in the total field angle $\theta$. (d) State trajectory (red solid line) on the Bloch sphere in the adiabatic frame, where the total field points constantly in the $\hat{z}$-direction (blue solid line). We have magnified the area of the sphere around the north pole.}
\label{fig:three_off}
\end{figure*}

As a last example, we find the pulse-sequence with total duration $T'=4.5\pi$ in the rescaled time $\tau$. Since $T'_2<T'<T'_3$, the pulse-sequence contains $m=3$ ``off" pulses, thus it has the form ``on-off-on-off-on-off-on", see Fig. \ref{fig:pulse3}. Working as in the previous case, Eq. (\ref{a_y}) becomes
\begin{widetext}
\begin{eqnarray}
\label{ay3}
a_{y,3}&=&\frac{1}{2}\mbox{Tr}(\sigma_yU)=\frac{1}{2}\mbox{Tr}(\sigma_yU_1W_2U_3W_2U_3W_2U_1)=\frac{1}{2}\mbox{Tr}(U_3W_2U_1\sigma_yU_1W_2U_3W_2)\nonumber\\
&=&in_y\big[\cos{(\tau_2/2)}\cos{\omega \tau_3}-n_z\sin{(\tau_2/2)}\sin{\omega \tau_3}\big]\big[\sin{\omega \tau_1}\cos{\tau_2}-n_z\sin{\tau_2}(1-\cos{\omega \tau_1})\big]+\nonumber\\
&&in_y\big[n_z\cos{(\tau_2/2)}\sin{\omega \tau_3}+\sin{(\tau_2/2)}(n_y^2+n_z^2\cos{\omega \tau_3})\big]\big[-\sin{\omega \tau_1}\sin{\tau_2}+n_z(1-\cos{\omega \tau_1})(1-\cos{\tau_2})\big]+\nonumber\\
&&in_y\big[\cos{\omega \tau_1}\cos{(\tau_2/2)}\sin{\omega \tau_3}-n_z\sin{(\tau_2/2)}(1-\cos{\omega \tau_1}\cos{\omega \tau_3})\big].
\end{eqnarray}
\end{widetext}
As before, we quickly check the above complicated formula. If we set $\tau_2=0$, thus the pulse-sequence degenerates to a constant ``on" pulse of duration $T'=2(\tau_1+\tau_3)$, then Eq. (\ref{ay3}) reduces consistently to $a_{y,3}=in_y\sin{\omega (\tau_1+\tau_3)}=0$. If we set $\tau_3=0$, so the pulse-sequence degenerates to a ``on-off-on" sequence where the ``off" pulse has duration $3\tau_2$, then Eq. (\ref{ay3}) reduces to the form (\ref{ay1}) with $3\tau_2$ in place of $\tau_2$. The final test is performed by equating to zero the duration of the middle ``off" pulse, in which case the pulse-sequence degenerates to a ``on-off-on-off-on" sequence where the middle ``on" pulse has duration $2\tau_3$. In order to account for this case correctly, in Eq. (\ref{ay3}) we need to set $\sin{(\tau_2/2)}=0$, $\cos{(\tau_2/2)}=1$, while keeping $\sin{\tau_2}, \cos{\tau_2}$. Then, it can be easily verified that this equation reduces to the form (\ref{ay2}) with $2\tau_3$ in place of $\tau_3$.

As in the previous example, using Eqs. (\ref{first_relation}), (\ref{second_relation}) in Eq. (\ref{ay3}), we can express $a_{y,3}$ as a function of duration $\tau_1$ and the amplitude $u$ of the pulse-sequence. For the total duration $T'=4.5\pi$ that we use here, the minimum value of $u$ for which the transcendental equation $a_{y,3}=0$ has a solution with respect to $\tau_1$ is found numerically to be $u=0.235698$. In Fig. \ref{fig:fidelity3} we plot the logarithmic error
\begin{equation}
\log_{10}{(1-F)}=\log_{10}{|b_2(T')|^2}=\log_{10}{|a_{y,3}|^2}
\end{equation}
as a function of $\tau_1$, for $u=0.235698$ and $T'=4.5\pi$. The observed resonance corresponds to the solution of the transcendental equation $a_{y,3}=0$.

\begin{figure}[t]
 \centering
		\begin{tabular}{c}
     	
      \subfigure[$\ $]{
	            \label{fig:detuning1}
	            \includegraphics[width=.7\linewidth]{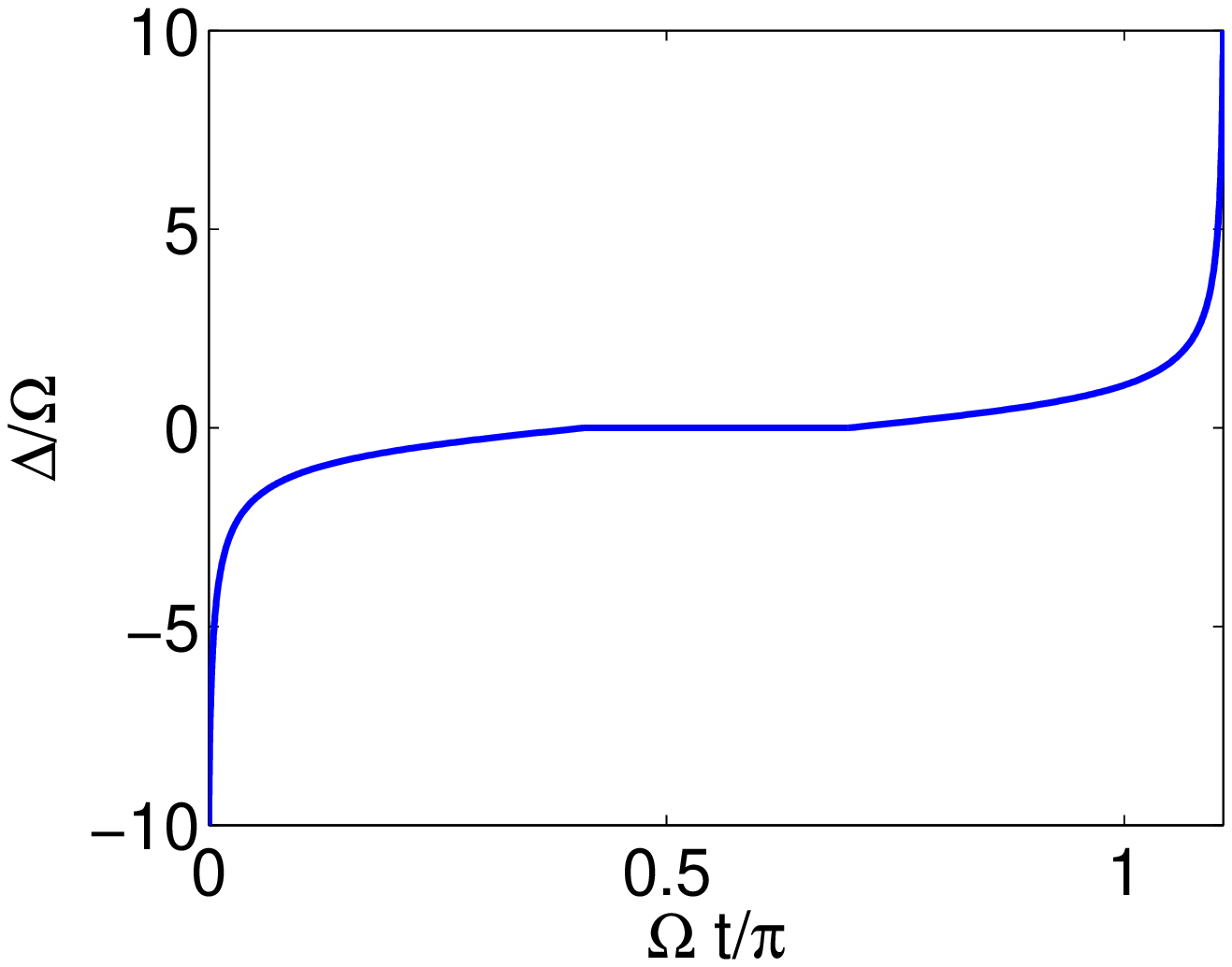}} \\
      \subfigure[$\ $]{
	            \label{fig:detuning2}
	            \includegraphics[width=.7\linewidth]{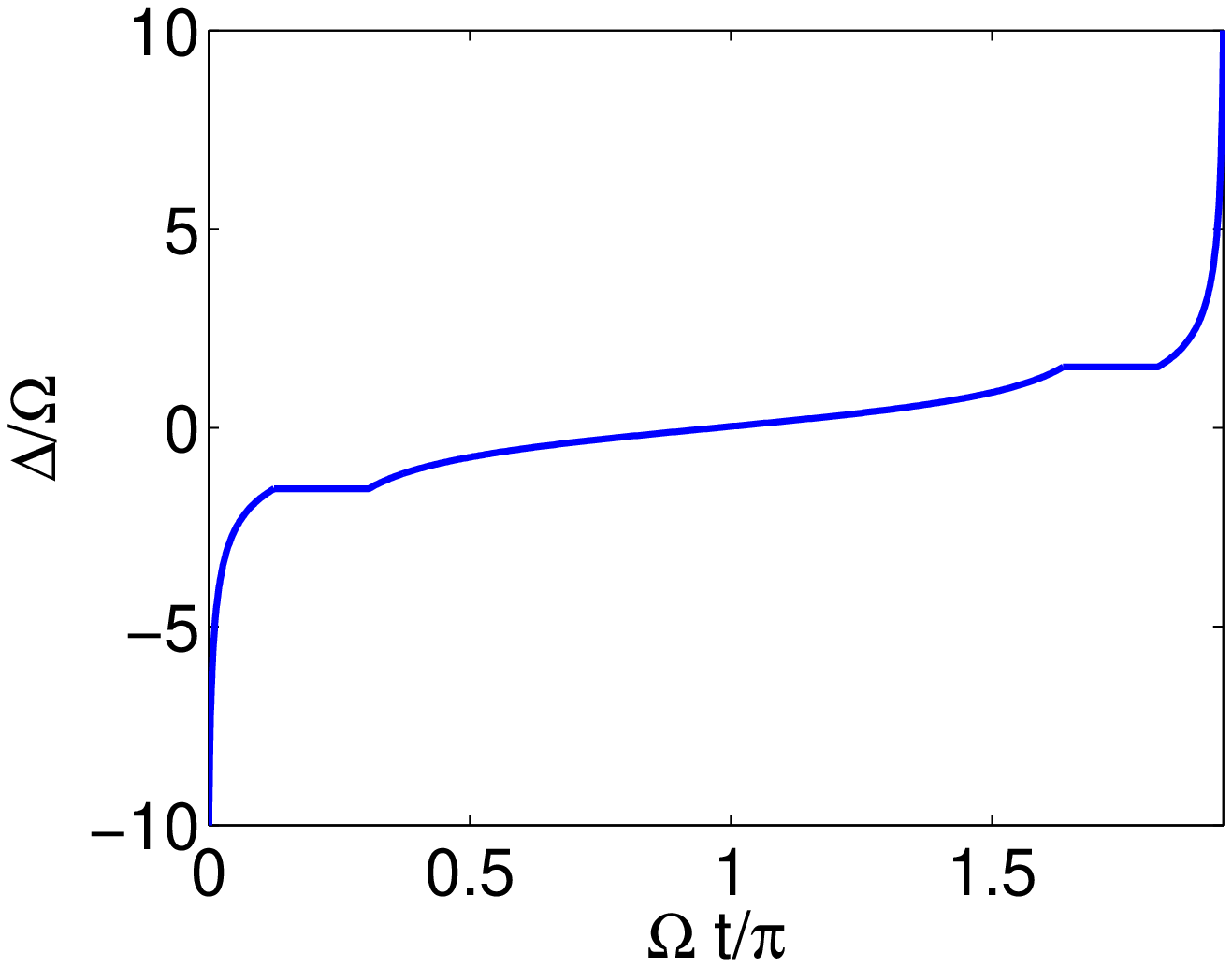}} \\
       \subfigure[$\ $]{
	            \label{fig:detuning3}
	            \includegraphics[width=.7\linewidth]{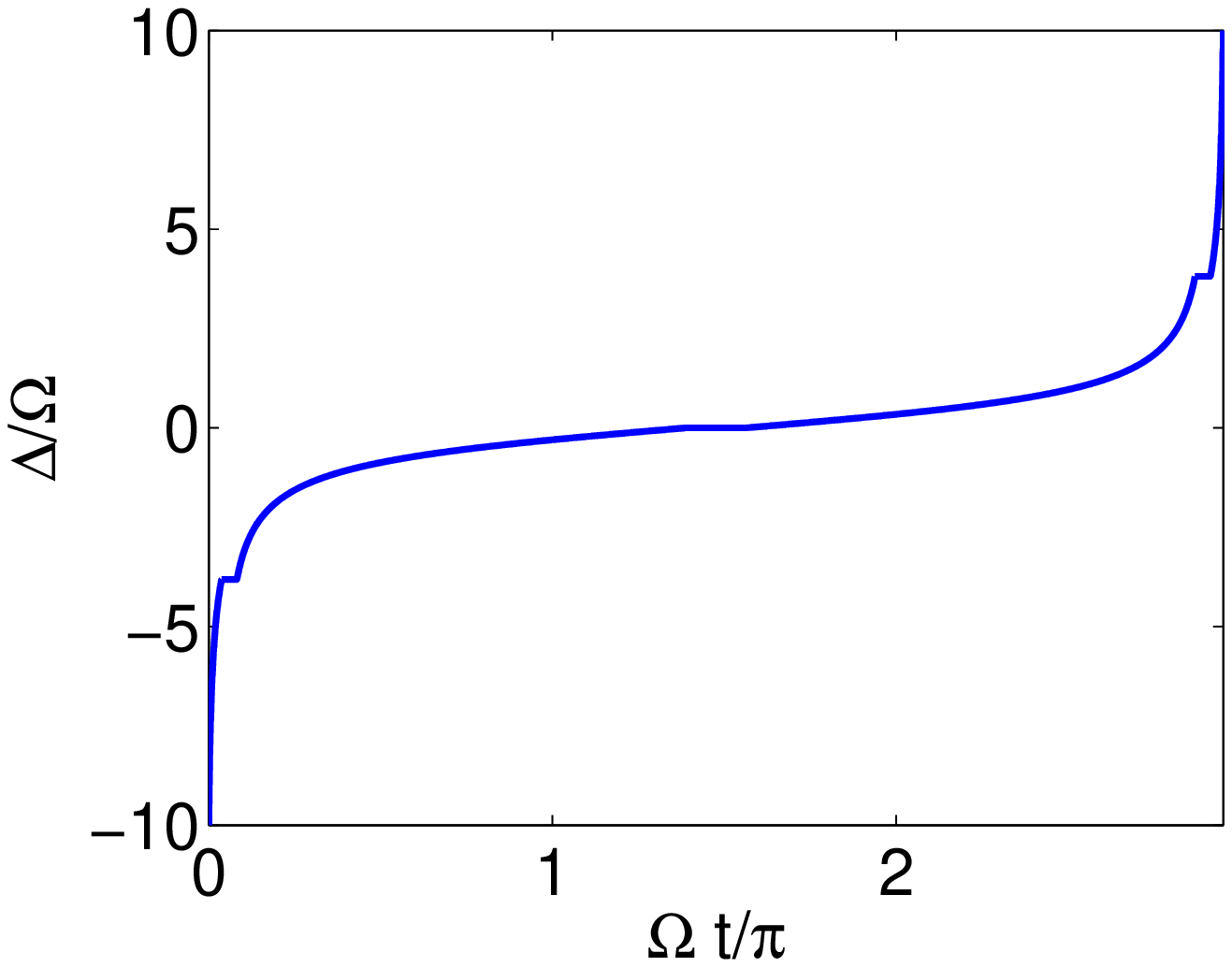}}
		\end{tabular}
\caption{(Color online) Detuning $\Delta(t)$ in the original time $t$ for the three presented examples: (a) on-off-on, (b) on-off-on-off-on, (c) on-off-on-off-on-off-on.}
\label{fig:detunings}
\end{figure}

In Fig. \ref{fig:pulse3} we plot the pulse sequence $u(\tau)$ corresponding to duration $T'=4.5\pi$ in the rescaled time $\tau$. In Fig. \ref{fig:theta3} we show the corresponding evolution of the total field angle $\theta(t)$ in the original time $t$, where the duration corresponding to $T'$ is found by integrating Eq. (\ref{rescaled_time}) to be $T\approx 2.952\pi/\Omega$. Observe that, although in the rescaled time all the ``off" pulses have the same duration, in the original time the middle ``off" pulse is longer. The reason is that for the middle pulse the coefficient $\sin{\theta}$ in Eq. (\ref{rescaled_time}) is larger than for the other two pulses. The detuning $\Delta(t)$ in the original time $t$ is displayed in Fig. \ref{fig:detuning3}. In Fig. \ref{fig:original3} we plot with red solid line the corresponding state trajectory on the Bloch sphere and in the original reference frame. The blue solid line on the meridian indicates the change in the total field angle $\theta$. Finally, in Fig. \ref{fig:adiabatic3} we plot the same trajectory (red solid line) but in the adiabatic frame, where the total field points constantly in the $\hat{z}$-direction (blue solid line). Observe the additional loop which can be hardly distinguished. This small loop in Fig. \ref{fig:adiabatic3} corresponds in Fig. \ref{fig:original3} to the approach to the meridian around $\theta=\pi/2$.

\section{Conclusion}

\label{sec:conclusion}

In this article, we derived novel ultrafast shortcuts for ARP in a two-level system with only longitudinal $z$-field control. In the adiabatic reference frame with appropriately rescaled time and using as control signal the derivative of the total field polar angle (with respect to rescaled time), we found composite pulses which achieve perfect fidelity for every duration larger than the limit $\pi/\Omega$, where $\Omega$ is the constant transverse $x$-field. The corresponding control $z$-field is a continuous function of the original time. The present work is expected to find applications in various tasks in quantum information processing, for example the design of a high fidelity controlled-phase gates, but also in other physical contexts where ARP is used.

\begin{acknowledgments}
The research is implemented through the Operational Program ``Human Resources Development, Education and Lifelong Learning'' and is co-financed by the European Union (European Social Fund) and Greek national funds (project E$\Delta$BM34, code MIS 5005825).
\end{acknowledgments}

\end{document}